\documentclass{article}
\usepackage{epsfig}
\usepackage{amsmath}
\usepackage{amsfonts}\tolerance=10000
\date{\today} 
 
\usepackage{graphicx}

\newcommand{\insertplot}[5]{\begin{figure}
 \hfill\hbox to 0.05in{\vbox to #5in{\vfill
 \inputplot{#1}{#4}{#5}}\hfill}
 \hfill\vspace{-.1in}
 \caption{#2}\label{#3}
 \end{figure}}
 \newcommand{\inputplot}[3]{
 \special{ps: plotfile #1}
}
\newcounter{fig}

\newcommand{\beq}{\begin{equation}}
\newcommand{\eeq}{\end{equation}}
\newcommand{\beqs}{\begin{eqnarray}}
\newcommand{\eeqs}{\end{eqnarray}}

\numberwithin{equation}{section}
\newcommand{\be}{\begin{equation}}
\newcommand{\ee}{\end{equation}}
\newcommand{\bea}{\begin{eqnarray}}
\newcommand{\eea}{\end{eqnarray}}

\usepackage{graphicx}

\begin{document}

\title{ Electric charge on the brane?} 
  
 \author{
 {\large Burkhard Kleihaus}$^{\dagger}$, {\large Jutta Kunz}$^{\dagger}$, 
 {\large Eugen Radu}$^{\ddagger \star }$ 
  and {\large  Daniel Senkbeil} $^{\dagger}$
\\ 
\\
$^{\dagger}${\small Institut f\"ur Physik, Universit\"at Oldenburg, Postfach 2503
D-26111 Oldenburg, Germany} 
\\
$^{\ddagger }${\small School of Theoretical Physics -- DIAS, 10 Burlington
Road, Dublin 4, Ireland }
\\
$^{\star}${\small  Department of Computer Science,
National University of Ireland Maynooth,
Maynooth,
Ireland}  
 }
 
 \maketitle 

\begin{abstract}
We consider black holes localized
on the brane in the Randall-Sundrum infinite braneworld model.
These configurations are static and charged with respect 
to a spherically symmetric, electric Maxwell field living on the brane.
We start by attempting to construct vacuum black holes, 
in which case our conclusions are in agreement with those of Yoshino
in
JHEP 0901:068, 2009
(arXiv:0812.0465). 
Although approximate solutions appear to exist 
for sufficiently small brane tension, 
these are likely only numerical artifacts.
The qualitative features of the configurations
in the presence of a brane U(1) electric field 
are similar to those in the vacuum case.
In particular, we find a systematic unnatural behaviour 
of the metric functions in the asymptotic region 
in the vicinity of the AdS horizon.
Our results are most naturally interpreted 
as evidence for the nonexistence of static, nonextremal 
charged black holes on the brane.
In contrast, extremal black holes are more likely to exist on the brane.
We determine their near-horizon form
by employing both analytical and numerical methods.
For any bulk dimension $d>4$, 
we find good agreement between the properties of large 
extremal black holes and
the predictions of general relativity, 
with calculable subleading corrections. 
\end{abstract}

\section{ Introduction}

The Randall-Sundrum (RS) infinite braneworld scenario \cite{Randall:1999vf} has been proposed as
a mechanism to explain the hierarchy between the TeV scale and the Planck scale,
and to realize four-dimensional gravity effectively on the 3-brane.
In this model, the observable universe is a 3-brane (domain wall) to which 
standard model fields are confined, 
while gravity can access the extra spatial dimensions.
The bulk metric is a locally Anti-de Sitter (AdS) spacetime 
satisfying the Einstein equations with negative cosmological constant. 
At low enough energy, perturbative Newtonian gravity
is recovered on the brane at distances large compared to the 
AdS length scale $\ell$.

The existence of black hole solutions in the RS model 
is an interesting open problem 
(see $e.g.$~Kanti \cite{Kanti:2009sz} for a recent review on this issue).
A priori, it is not clear why such black hole solutions should not exist,
and the existence of bulk black hole solutions would imply
that there are also black hole solutions on the brane.
Indeed, such exact black hole solutions were found 
by Emparan et al.~\cite{Emparan:1999wa}
for the dimension $d=4$ of the bulk spacetime, 
where the static black hole is localized on a 2-brane.
The construction of such black hole solutions in \cite{Emparan:1999wa} relied
on the existence of a special class of solutions
of the $d=4$ Einstein gravity -- 
the so-called C-metric \cite{Plebanski:1976gy}, 
and the brane is a suitable slice in this spacetime.

However, in the absence of a $d>4$ generalization of the C-metric,
the existence of higher dimensional brane world black holes is 
controversial. 
Although Dadhich et al.~\cite{Dadhich:2000am} proposed 
an analytic solution for brane black holes in the $d=5$ RS scenario, 
this result was obtained subject to approximations 
whose validity is not clear. 
In that approach, the braneworld Einstein equations \cite{Shiromizu:1999wj}
were solved assuming a particular form for the bulk Weyl tensor on the wall.
Interestingly, this resulted in a Reissner-Nordstr\"om (RN) geometry, 
although there is no brane electric charge.
However, in this case  only the induced metric on the brane was found, 
without solving the bulk equations of motion. 
Thus the solution proposed in \cite{Dadhich:2000am} for a brane black hole 
in the RS scenario is not really satisfactory. 
 
Hence, for bulk dimensions $d>4$, 
one has to rely on numerical methods 
to address the existence of black holes in the RS model.
So far, the only results supporting the existence of such solutions 
on a nonperturbative level are those by 
Kudoh et al.~\cite{Kudoh:2003xz,Kudoh:2003vg}.
They were obtained within an in principle rigorous approach,
$i.e.$ by solving the full set of equations in the bulk 
with suitable boundary conditions on the brane.
Reporting numerical results for spacetime dimensions $d=5$ and $6$,
Kudoh et al.~\cite{Kudoh:2003xz,Kudoh:2003vg}
presented localized black holes
whose horizon radius was much smaller than the AdS length scale.
In contrast, for larger black holes their numerical accuracy decreased 
and convergence was lost within their numerical scheme.

Moreover, the recent numerical work of Yoshino \cite{Yoshino:2008rx} 
(while finding agreement with earlier perturbative \cite{Karasik:2003tx}
and numerical \cite{Kudoh:2003xz,Kudoh:2003vg} results)
also failed to find such large black hole configurations.
Performing a careful error analysis in terms of
systematic and nonsystematic errors, Yoshino \cite{Yoshino:2008rx}
concluded
that the positive results in \cite{Kudoh:2003xz,Kudoh:2003vg} 
were nothing but numerical artifacts, mainly due to an
inappropriate treatment of the outer boundary of the integration region.

In fact, Emparan et al.~\cite{Emparan:2002px} 
and also Tanaka \cite{Tanaka:2002rb} had earlier put forward
a number of theoretical arguments against the existence 
of $d>4$ {\it static} black holes on the brane in the RS model, 
which were mainly based on a version of the AdS/CFT correspondence.
According to the conjecture by Emparan et al.~\cite{Emparan:2002px}, 
such bulk black holes would necessarily be time-dependent, 
since their duals would describe quantum corrected 
black holes in a $d-1$ dimensional braneworld. 
Counter-arguments were, however, 
given by Fitzpatrick et al.~\cite{Fitzpatrick:2006cd}.
Further results on  black holes localized
on the brane  
can be found in \cite{Anderson:2004md}-\cite{Kanti:2003uv}.
 
Motivated by these conflicting arguments and results, 
we have performed an independent investigation of the issue of 
the numerical construction of $d>4$ braneworld black holes.
Similar to the previous work by Kudoh et al.~\cite{Kudoh:2003xz,Kudoh:2003vg} 
and Yoshino \cite{Yoshino:2008rx}, we have solved the bulk
Einstein equations with Israel junction conditions on the brane,
employing different numerical methods from those in 
\cite{Kudoh:2003xz,Kudoh:2003vg,Yoshino:2008rx}.
In particular, we have used a compactified radial coordinate 
which in principle avoids the problems 
associated with the position of the outer boundary of integration 
for the radial coordinate. 

The results we have found for bulk dimensions $d=5$ and $6$ 
support the claim of Yoshino \cite{Yoshino:2008rx} that
{\it `a solution sequence of a static black hole
on an asymptotically flat brane 
that is reduced to the Schwarzschild black hole in the zero
tension limit is unlikely to exist'}.
In particular, we have noticed a systematic unnatural 
behaviour of the metric functions 
for large values of the radial coordinate, 
$i.e.$~close to the AdS horizon. 
This behaviour seems to be at the origin of the loss of 
numerical convergence for large black holes.

Obviously, it is interesting to examine how generic this behaviour is,
and whether it can be circumvented in more general cases.
Perhaps the simplest more general case to be studied 
corresponds to black hole solutions, 
which are $electrically$ charged with respect to 
a Maxwell field living on the brane.
This type of solutions was considered 
by Chamblin et al.~\cite{Chamblin:2000ra},
following a different approach, however.
There the ``initial'' data on the brane was prescribed,
and then it was evolved in the spacelike direction
transverse to the brane, 
by solving the bulk equations numerically.
The results in \cite{Chamblin:2000ra} 
show the occurrence of pathological features in the bulk 
for any initial data.
However,  Chamblin et al.~\cite{Chamblin:2000ra}
postulated a special restricted form of the braneworld metric, 
with a single essential function;
moreover, the numerical integration employed 
a relatively small cutoff radius.

Our present approach is rather different from \cite{Chamblin:2000ra}, 
since we attempt to directly solve the 
bulk vacuum Einstein equations with suitable boundary conditions.
The bulk theory is the same as for the uncharged case,
the electric charge entering the problem via 
the Israel junction conditions on the brane.
Restricting to static, nonextremal configurations in an AdS$_5$ bulk, 
our results show that all pathologies present for vacuum black holes 
occur also in this case. 
Therefore, we conclude that the existence of 
static, charged, nonextremal solutions is unlikely, as well.

Turning next to static, extremal,
electrically charged black holes in the RS brane world scenario,
however,
we anticipate that such solutions are more likely to exist,
since a number of arguments put forward against the existence
of static black holes do not apply when the Hawking temperature vanishes.
Here we investigate the near-horizon structure of extremal black hole solutions
with electric charge on the brane, 
without attempting to construct the full configurations.
This restriction leads to a system of coupled nonlinear 
ordinary differential equations, which are solved
numerically within a nonperturbative approach for 
several dimensions $d\geq 5$ of the bulk.

For a five-dimensional AdS bulk, 
this problem has been considered 
by Kaus and Reall \cite{Kaus:2009cg}. 
In this work we generalize their results for 
any $d\geq 5$ dimensions, and show that
for large black holes, there is a good agreement 
for the induced metric on the brane and for the entropy 
with the predictions of general relativity (GR),
with calculable subleading corrections.

The paper is organized as follows:
in the next Section we present our results 
for nonextremal black holes in the RS brane world model,
employing a nonperturbative approach,
by directly solving a set of three nonlinear partial differential equations 
with suitable boundary conditions.
In Section 3 we consider extremal black holes 
that are charged with respect to a purely electric Maxwell field
on the brane and determine their near-horizon form.
We give our conclusions and remarks in the final Section.
The Appendix contains a
discussion of some technical aspects involved in our numerical investigation
of nonextremal brane world black holes.

\section{Nonextremal configurations}

\subsection{The problem}
We consider the  RS braneworld model, 
with a $d$-dimensional bulk spacetime
and a single $(d-1)$-dimensional brane with positive tension in it.
Also, we impose $Z_2$-symmetry about the brane, 
which is assumed to be asymptotically flat.
The bulk matter is merely a negative cosmological constant,
and the brane tension and the matter localized on the brane 
are treated in a distributional sense.

The action of this model is  
\begin{eqnarray}
S = \frac{1}{16 \pi G_{d}}\int_{\cal M} d^dx\sqrt{-g}  \left( R -2 \Lambda\right) 
  +\int_{\rm brane} d^{d-1}x \sqrt{-h} 
  \left(\frac{1}{8 \pi G_{d}}K-\sigma -\frac{1}{16 \pi G_{d-1}}F_{\mu\nu}F^{\mu\nu}\right), 
\end{eqnarray}
where ${\cal M}$ is the bulk spacetime, 
$\Lambda=-(d-2)/(d-1)/(2\ell^2)$ is the bulk cosmological constant
and $K_{\mu\nu}$ is the projection
of the extrinsic curvature of the brane hyper-surface with induced metric
$h_{\mu\nu}$. 
Also $G_{d}$ is Newton's constant in $d$-spacetime
dimensions;
$\sigma$ and $F=dA$ are the brane tension and 
the field strength of the Maxwell field on the brane, respectively. 

From the above action, we obtain the $d$-dimensional Einstein equation in the 
bulk  
\begin{eqnarray}
\label{eqs-einstein}
R_{ij}-\frac{1}{2}Rg_{ij}+\Lambda g_{ij}=0. 
\end{eqnarray}
For the RS infinite braneworld scenario, 
the Israel junction conditions on the 
brane are given by~\cite{Israel:1966rt}
\begin{eqnarray}
 \label{eqs-israeljci}
 K_{\mu\nu}-K h_{\mu\nu}  = 4 \pi G_d \left( -\sigma h_{\mu \nu}+\frac{1}{4 \pi G_{d-1}} t_{\mu\nu} \right ), 
\end{eqnarray}
where $t_{\mu\nu}$ is the energy-momentum tensor of the matter fields on the brane.
For a U(1) field, the  expression of $ t_{\mu\nu}$ is
\begin{eqnarray}
\label{tik-U1}
t_{\mu\nu} =  F_{\mu \alpha}{F_\nu}^\alpha-\frac{1}{4}F_{\alpha \beta}F^{\alpha \beta} h_{\mu\nu}.
\end{eqnarray}
We shall set the brane tension to the RS value
$\sigma=\frac{1}{4\pi G_d} \frac{d-2}{\ell}$, while $G_{d-1}=\frac{d-3}{2\ell} G_d$. 
This simplifies eq.~(\ref{eqs-israeljci}) to
 \begin{eqnarray}
 \label{eqs-israeljc}
 K_{\mu\nu}-K h_{\mu\nu}  =   -\frac{d-2}{\ell} h_{\mu \nu} 
 +\frac{2\ell}{d-3} t_{\mu\nu} , 
\end{eqnarray}
which is the form used in what follows.
The brane U(1) field is a solution of the Maxwell equations
 \begin{eqnarray}
 \label{eqs-maxwell}
\nabla_{\mu} F^{\mu \nu}=0, 
\end{eqnarray}
for a metric background given by $h_{\mu \nu}$.

\subsubsection{The ansatz and the equations}
The metric ansatz employed here essentially corresponds to the one used 
in the previous studies \cite{Kudoh:2003xz,Kudoh:2003vg,Yoshino:2008rx}.
The black hole metric is spherically symmetric on the brane 
and axisymmetric in the bulk spacetime, with line element 
\begin{eqnarray}
\label{metric-ansatz}
ds^2=\frac{1}{z^2(r,\chi)}
\left(
e^{2B(r,\chi)}(\frac{dr^2}{F(r)}+r^2 d \chi^2)
+e^{2 C(r,\chi)}r^2 \sin^2 \chi d\Omega_{d-3}^2
-e^{2A(r,\chi)}F(r) dt^2
\right),
\end{eqnarray}
which is parametrized in terms of two background functions
 \begin{eqnarray}
 \label{bckgf}
F(r)=1-\left( \frac{r_0}{r}\right )^{d-3}, ~~
z(r,\chi)=1+\frac{r}{\ell}\cos\chi,
\end{eqnarray}
and three unknown metric functions $A(r,\chi)$, $B(r,\chi)$ and $C(r,\chi)$.
In the above relations, $r_0$ is a positive constant 
and $d\Omega_{d-3}^2$ is the metric on the $(d-3)$-sphere.
The background functions $F(r)$ and $z(r,\chi)$ 
have been introduced such that two important limits 
of the general solution are already contained within the 
ansatz (\ref{metric-ansatz}).  
For $\ell\to \infty$ one finds the well-known Schwarzschild-Tangherlini
black hole, expressed in the usual Schwarzschild 
coordinates\footnote{Refs.~\cite{Kudoh:2003xz,Kudoh:2003vg} and 
\cite{Yoshino:2008rx} preferred to describe
the Schwarzschild black hole in isotropic coordinates, 
which was also our initial choice.
However, we realized that a Schwarzschild coordinate system together with 
the coordinate transformation (\ref{coord-transf})
improved the quality of the numerical calculations for $d=5$.}. 
Another limit of interest is $r_0=0$, 
in which case one recovers the original RS model, 
$i.e.$ a part of AdS$_d$ spacetime expressed
in Poincar\'e coordinates (and $A=B=C=0$ in both limits).

The event horizon is supposed to reside at a surface of constant radial 
coordinate $r = r_0$ and characterized by the condition $F(r_0) = 0$, 
while the brane is located at $\chi=\pi/2$. 
Then the coordinate range 
considered is $r_0\leq r<\infty$ and $0\leq \chi\leq \pi/2$. 
Therefore the coordinates in eq.~(\ref{metric-ansatz}) 
have a rectangular boundary and thus are suitable for 
the numerical methods employed.
 
The induced metric on the brane has line 
element\footnote{Note the presence of
three distinct metric functions in eq.~(\ref{ans2}).
In contrast, in GR only two metric functions are present, with 
the usual choice  $g_{rr}=1/N(r)$, $g_{\Omega \Omega}=r^2$ and
$g_{tt}=N(r)\sigma^2(r)$.}
\begin{eqnarray}
\label{ans2}
d\sigma^2= g_{rr}(r) dr^2+g_{\Omega \Omega}(r)d\Omega_{d-3}^2+g_{tt}(r)dt^2,
 \end{eqnarray}
 with 
\begin{eqnarray}
 g_{rr}(r)= \frac{e^{2B(r,\pi/2)}}{F(r)},~~~g_{\Omega \Omega}(r)=e^{2C(r,\pi/2)}r^2,~~g_{tt}(r)= -e^{2A(r,\pi/2)}F(r),
 \end{eqnarray}
$i.e.$ it describes a static, spherically symmetric black hole spacetime
in $d-1$ dimensions.

The  equations satisfied by the functions $A,~B $ and $C$ 
are found by using 
a suitable combination 
of the Einstein equations, 
$G_t^t+\Lambda=0,~G_r^r+G_\chi^\chi+2\Lambda=0$ and $G_{\phi}^{\phi}+\Lambda=0$
(where $\phi$ denotes an angle of
the $d - 3$ dimensional sphere) and read\footnote{One can see that 
the case $d=4$ is special, since a number of terms vanish in this case.
However, the structure of the equations is the same for any $d>4$.}
\begin{eqnarray}
\nonumber
A''+\frac{1}{r^2F}{\ddot A}
+(\frac{d-2}{r}+\frac{3F'}{2F})(A'-\frac{\cos \chi}{\ell z})
+\frac{(d-3)F'}{2F}(C'-\frac{\cos \chi}{\ell z})
+(A'-\frac{\cos \chi}{\ell z})^2
\\
\label{neq-A}
+(d-3)(A'-\frac{\cos \chi}{\ell z})(C'-\frac{\cos \chi}{\ell z})
+\frac{(d-3)\cot \chi}{r^2 F}(\dot A+\frac{r \sin \chi}{\ell z})
+\frac{1}{r^2 F}(\dot A+\frac{r \sin \chi}{\ell z})^2
\\
\nonumber
+\frac{(d-3)}{r^2 F}(\dot A+\frac{r \sin \chi}{\ell z})(\dot C+\frac{r \sin \chi}{\ell z})
-\frac{(d-1)e^{2B}}{\ell^2 F z^2}
+\frac{1}{r F}\frac{(r+\ell \cos \chi)}{\ell^2 z^2}
+\frac{\cos^2 \chi}{\ell^2 z^2}=0,
\end{eqnarray}
\begin{eqnarray}
\nonumber
B''+\frac{1}{r^2F}{\ddot B} 
-\frac{(d-3)}{r}\left(A'-\frac{\cos \chi}{\ell z}+(d-4)(C'-\frac{\cos \chi}{\ell z})\right)
+\frac{1}{r}(B'-\frac{\cos \chi}{\ell z})
+\frac{F'}{2F}(B'-\frac{\cos \chi}{\ell z})
\\
\nonumber
-\frac{(d-3)F'}{2F}(C'-\frac{\cos \chi}{\ell z})
-(d-3)(A'-\frac{\cos \chi}{\ell z})(C'-\frac{\cos \chi}{\ell z})
-\frac{1}{2}(d-3)(d-4)(C'-\frac{\cos \chi}{\ell z})^2
\\
\label{neq-B}
-(d-3)\frac{\cot\chi}{r^2 F}\left(\dot A+\frac{r \sin \chi}{\ell z}
+(d-4)(\dot C+\frac{r \sin \chi}{\ell z})\right)
-\frac{(d-3)}{r^2F}(\dot A+\frac{r \sin \chi}{\ell z})(\dot C+\frac{r \sin \chi}{\ell z})
\\
\nonumber
-\frac{(d-3)(d-4)}{2r^2F}(\dot C+\frac{r \sin \chi}{\ell z})^2
+(d-3)(d-4)\frac{e^{2(B-C)}}{2\sin^2\chi F r^2}
+\frac{(d-1)(d-4)e^{2B}}{2\ell^2 F z^2}
\\
\nonumber
-\frac{(d-3)(d-4)}{2r^2}(1+\frac{\cot^2\chi}{F})
-\frac{(d-4)F'}{2r F}+\frac{1}{\ell^2 z^2}(\cos^2 \chi+\frac{r+\ell\cos \chi}{r F})=0,
\end{eqnarray}
\begin{eqnarray}
\nonumber
C''+\frac{1}{r^2 F}{\ddot C} 
+(d-3)(C'-\frac{\cos \chi}{\ell z})^2
+\frac{1}{r}(A'-\frac{\cos \chi}{\ell z})
+\frac{2d-5}{r}(C'-\frac{\cos \chi}{\ell z})
\\
\nonumber
+\frac{F'}{F}(C'-\frac{\cos \chi}{\ell z})
+(A'-\frac{\cos \chi}{\ell z})(C'-\frac{\cos \chi}{\ell z})
+\frac{(d-3)}{r^2 F}(\dot C+\frac{r \sin \chi}{\ell z})^2
+\frac{\cot \chi}{r^2 F}(\dot A+\frac{r \sin \chi}{\ell z})
\\
\label{neq-C} 
+\frac{2(d-3)\cot \chi}{r^2 F}(\dot C+\frac{r \sin \chi}{\ell z})
+\frac{1}{r^2 F}(\dot A+\frac{r \sin \chi}{\ell z})(\dot C+\frac{r \sin \chi}{\ell z})
-(d-4)\frac{e^{2(B-C)}}{r^2 F \sin^2 \chi}
\\
\nonumber
-(d-1)\frac{e^{2B}}{\ell^2 F z^2}
+\frac{r+\ell \cos \chi}{F \ell^2 r z^2}
+\frac{d-3}{r^2}
+\frac{\cos^2 \chi}{\ell^2 z^2}
+\frac{1}{r^2 F}((d-4)\cot^2\chi-1)
+\frac{F'}{rF}=0,
\end{eqnarray} 
where a  prime denotes the derivative with respect to the radial variable $r$ 
and a dot denotes the derivative with respect to the angular variable $\chi$.
The remaining equations $G_\chi^r =0,~G_r^r-G_\chi^\chi  =0$
yield two constraints. Following \cite{Wiseman:2002zc}, we note that
setting $G_t^t+\Lambda=0,~G_r^r+G_\chi^\chi+2\Lambda=0$, $G_{\phi}^{\phi}+\Lambda=0$
in $\nabla_\mu G^{\mu r} =0$ and $\nabla_\mu G^{\mu \chi}=0$, 
we obtain Cauchy-Riemann relations for $G_\chi^r$ and $G_r^r-G_\chi^\chi$.
Thus the weighted constraints satisfy Laplace equations, and the constraints 
are fulfilled, when one of them is satisfied on the boundary and the other 
at a single point
\cite{Wiseman:2002zc}.

For completeness, we here present also the expressions 
for the Hawking temperature $T_H$ 
and the event horizon area $A_H^{(d)}$ of a bulk black hole,
\begin{eqnarray}
\label{TH}
T_H=e^{A(0,\chi)-B(0,\chi)}\frac{(d-3)}{4\pi r_0}, ~~~~
A_H^{(d)}= V_{d-3}r_0^{d-2} \int_0^{\pi/2}e^{B(0,\chi)+(d-3)C(0,\chi)}\frac{\sin^{d-3}\chi}
{\left(1+\frac{r_0}{\ell}\cos \chi\right)^{d-2}} d\chi,
\end{eqnarray}
where $V_{d-3}$ is the area of the unit $S^{d-3}$ sphere. 
(The Einstein equation $G_\chi^r =0$ implies that
$e^{A -B }$ is indeed constant on the horizon.)
The associated black hole on the brane would have the same Hawking temperature 
as the bulk solution, while its event horizon area would be
\begin{eqnarray}
\label{AH-brane}
A_H^{(d-1)} =V_{d-3} r_0^{d-3} e^{(d-3)C(0,\pi/2)}.
 \end{eqnarray}

In practice, we have found it convenient to introduce 
the radial coordinate\footnote{This change of the radial coordinate
proved useful before in the numerical study of non-uniform black string solutions 
\cite{Kleihaus:2006ee}, which is
an axisymmetric problem with some similarities to the problem studied here.}
\begin{eqnarray}
\label{coord-transf}
\rho=\sqrt{r^2-r_0^2},
\end{eqnarray}
such that the horizon would reside at $\rho=0$, 
which gives a simpler set of boundary conditions for the black hole horizon. 
The transformation then results in a new form of the general 
ansatz (\ref{metric-ansatz}) with 
\begin{eqnarray}
\label{metric-ansatz1}
ds^2=\frac{1}{z^2(\rho,\chi)}
\left(
e^{2B(\rho,\chi)}(\frac{d\rho^2}{F_1(\rho)}+(\rho^2+r_0^2) d \chi^2)
+e^{2 C(\rho,\chi)}(\rho^2+r_0^2) \sin^2 \chi d\Omega_{d-3}^2
-e^{2A(\rho,\chi)}F_2(\rho) dt^2
\right),~~{~}
\end{eqnarray}
where $F_1(\rho)=F(r(\rho))(\rho^2+r_0^2)/\rho^2$ and $F_2(\rho)=F(r(\rho))$.
(Note that $F_1(\rho)\to (d-3)/2+O(\rho^2)$, 
$F_2(\rho)\to (d-3)\rho^2/(r_0^2)+O(\rho^4)$ as $\rho\to 0$.)
The  temperature  and the horizon area still given by (\ref{TH}), (\ref{AH-brane}).

\subsubsection{U(1) field on the brane and boundary conditions}
Perhaps the simplest example of nonvacuum solutions in the RS 
infinite braneworld scenario is provided by 
a Maxwell field confined to the brane.
In this work we shall restrict to a static, spherically symmetric, 
purely electric field with U(1) potential
\begin{eqnarray}
\label{V-a}
A=V(\rho) dt.
\end{eqnarray}
Thus the  field strength tensor is $F=\frac{dV}{d\rho} d\rho \wedge dt$.
The Maxwell equations (\ref{eqs-maxwell}) imply the existence of the first integral
\begin{eqnarray}
\label{V}
V(\rho)=Q\int  d \rho \frac{\rho}{(\rho^2+r_0^2)^{\frac{d-2}{2}}}e^{A(\rho,\pi/2)+B(\rho,\pi/2)-(d-3)C(\rho,\pi/2)},
\end{eqnarray}
where $Q$ is an integration constant that fixes the electric charge on the brane.

The numerical solution of the equations is pursued
subject to the following set of boundary conditions
\begin{eqnarray}
\label{bc1} 
\partial_\rho A\big|_{ \rho=0}=\partial_\rho B\big|_{ \rho=0}=\partial_\rho C\big|_{ \rho=0}=0,
\end{eqnarray}
on the black hole horizon,
\begin{eqnarray}
\label{bc2} 
 A\big|_{\rho=\infty}= B\big|_{ \rho=\infty}=  C\big|_{ \rho=\infty}=0,
\end{eqnarray}
at infinity, and
\begin{eqnarray}
\label{bc3} 
\partial_\chi A\big|_{\chi=0}=\partial_\chi  B\big|_{\chi=0}=\partial_\chi C\big|_{\chi=0}=0,
\end{eqnarray}
on the symmetry axis.
Regularity at $\chi=0$ further requires that $B|_{\chi=0} =C|_{\chi=0}$. 
This condition can be implemented by
working with the new function $\bar C=C-B$.
The Israel junction conditions (\ref{eqs-israeljc})  
together with the expression (\ref{V}) for the U(1) potential
lead to the following boundary conditions on the brane 
 \begin{eqnarray}
\label{bc-brane1} 
\nonumber
&\partial_\chi A\big|_{\chi=\pi/2}= \partial_\chi B\big|_{\chi=\pi/2}
=\frac{\sqrt{\rho^2+r_0^2}}{\ell}(e^{B(\rho, \pi/2)}-1)
+\frac{2d-7}{(d-2)(d-3)}e^{B(\rho, \pi/2)-2(d-3)C(\rho, \pi/2)}
\frac{\ell}{(\rho^2+r_0^2)^{(2d-7)/2}}Q^2,
\end{eqnarray}
 \begin{eqnarray}
\label{bc-brane2} 
\partial_\chi C\big|_{\chi=\pi/2}=\frac{\sqrt{\rho^2+r_0^2}}{\ell}(e^{B(\rho, \pi/2)}-1)
-\frac{3}{(d-2)(d-3)}e^{B(\rho, \pi/2)-2(d-3)C(\rho, \pi/2)}
\frac{\ell}{(\rho^2+r_0^2)^{(2d-7)/2}}Q^2,
\end{eqnarray} 
where $Q=0$ corresponds to the vacuum limit.

\subsection{Numerical results}

The results exhibited in this Section concern the physically most interesting 
case $d=5$ ($i.e.$ a 3-brane).
However, we have observed a similar behaviour  
when considering vacuum black holes for a 4-brane instead.

The numerical calculations have been based on the Newton-Raphson method
and been performed with help of the program FIDISOL/CADSOL \cite{schoen},
which also provides an error estimate for each unknown function.
Different from previous work on this problem, 
in our approach the boundary conditions (\ref{bc2})
are really imposed at $\rho=\infty$.
This is achieved by employing a compactified radial coordinate 
$x=\rho/(1+\rho)$ which maps $\rho=\infty$ to the finite value $x=1$.
Further details on the numerical method are given in the Appendix.
 
The first relevant input parameter for our problem is the dimensionless ratio   
\begin{eqnarray}
\label{rel} 
L=\frac{r_0}{\ell}.
\end{eqnarray} 
Without any loss of generality, by using a simple rescaling of $\rho$, 
one can set $r_0$ to take a fixed value,
and then vary the AdS length scale $\ell$. 
The results reported in this Section correspond to the choice $r_0=1$, 
although similar results have been found for other values of $r_0$. 
In the charged case, there is a second input parameter of the problem, 
which is the ratio  
\begin{eqnarray}
\label{rel2} 
q=\frac{Q}{r_0^{d-3}}.
\end{eqnarray}

\subsubsection{The vacuum case ($Q=0$)}

Similar to Kudoh et al.~\cite{Kudoh:2003xz,Kudoh:2003vg}
and Yoshino \cite{Yoshino:2008rx}, we have first applied
the numerical procedure to the above stated set of equations
and boundary conditions, choosing a large value of the AdS length scale
$\ell$ (typically $L \simeq 10^{-4}- 10^{-3}$) 
and employing the initial guess\footnote{We
note that for $L\neq 0$, $A=B=C=0$ is {\it not} 
a solution of the field equations unless $r_0=0$.} 
$A=B=C=0$.
\begin{figure}[t!]
\setlength{\unitlength}{1cm}
\setlength{\unitlength}{1cm}
\begin{picture}(15,18)
\put(-1,0){\epsfig{file=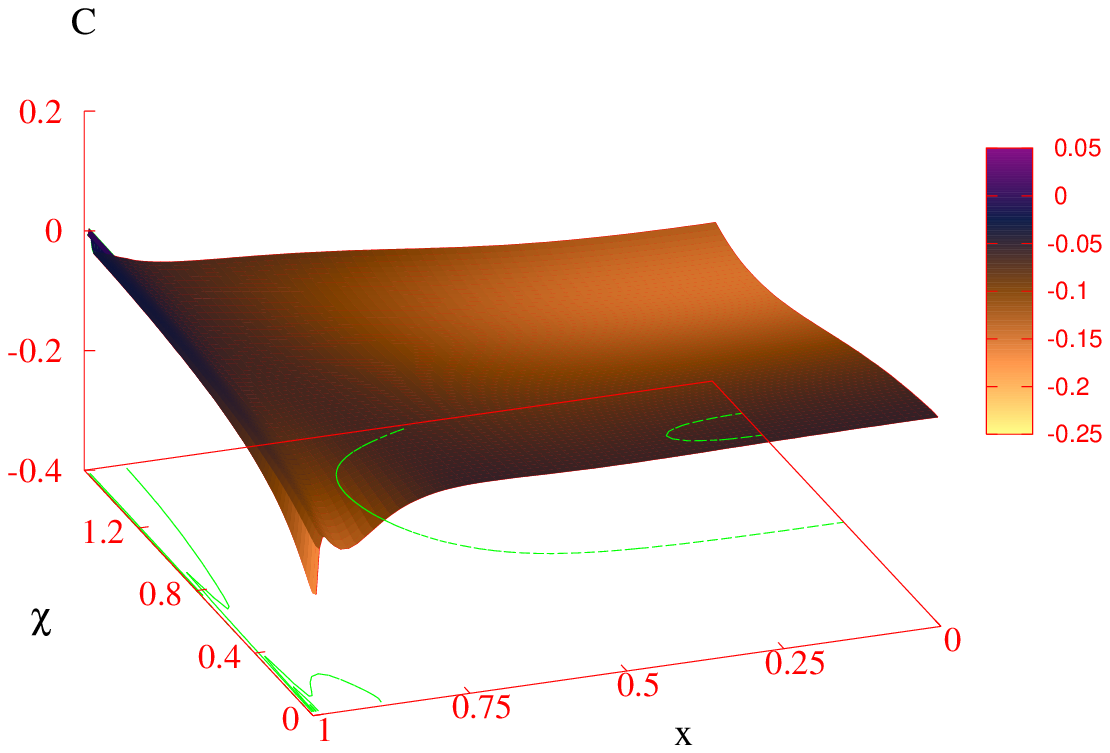,width=8cm}}
\put(7,0){\epsfig{file=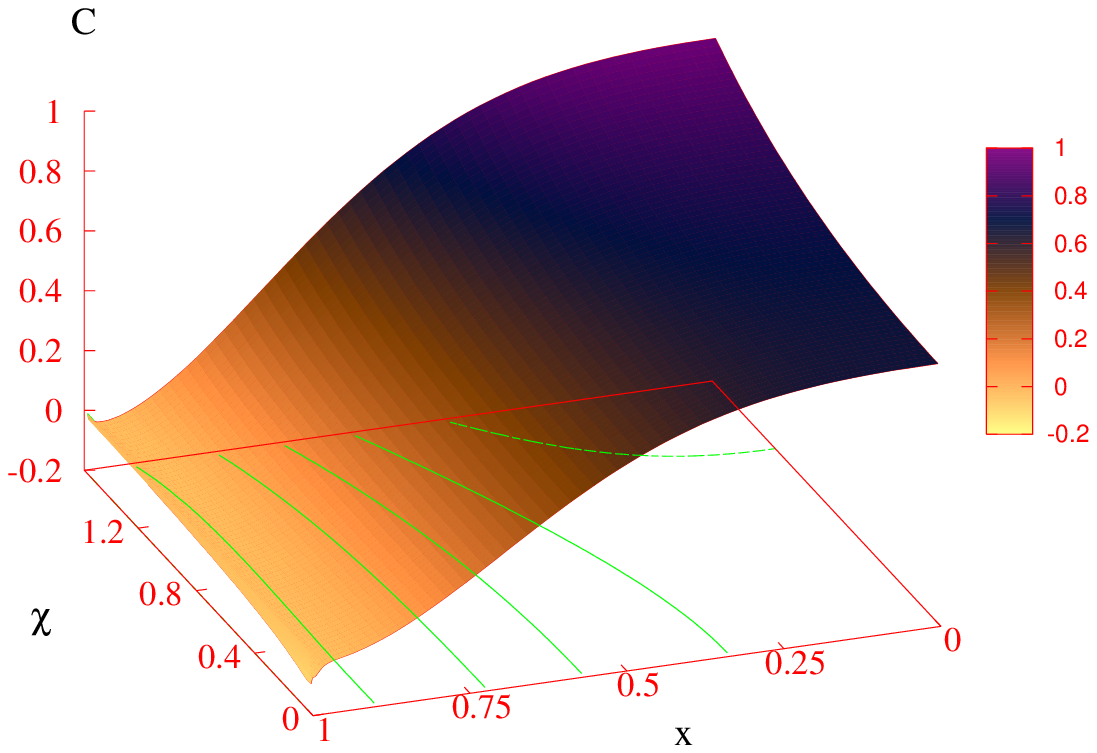,width=8cm}}
\put(-1,6){\epsfig{file=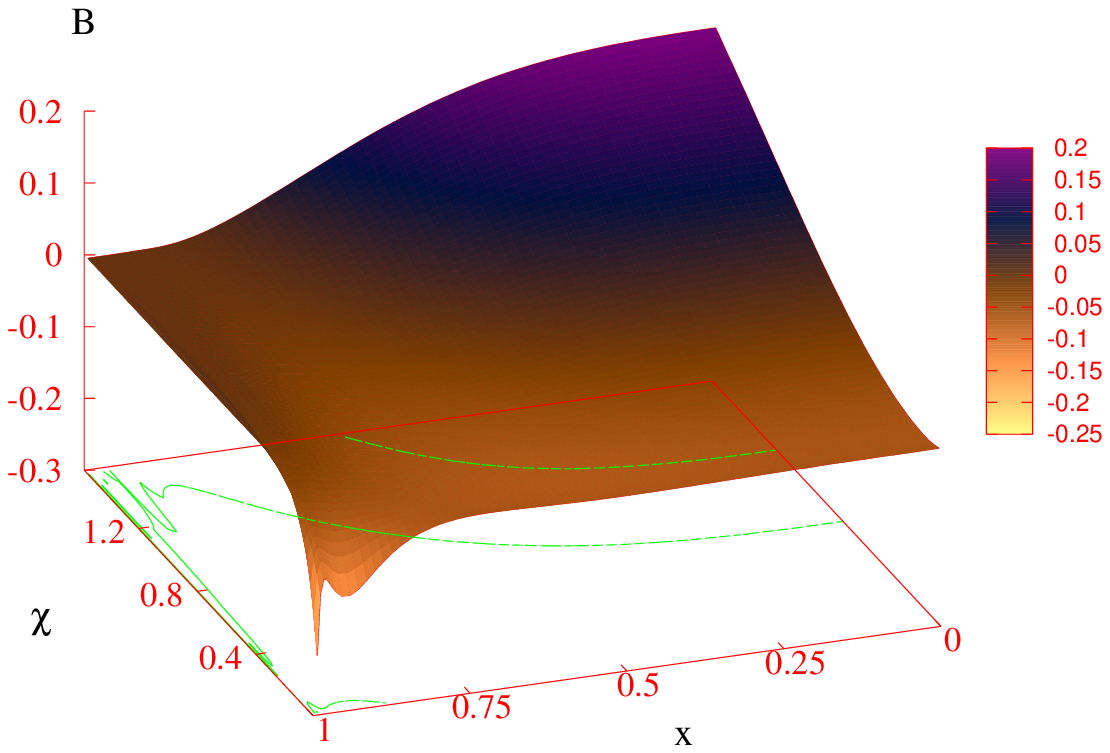,width=8cm}}
\put(7,6){\epsfig{file=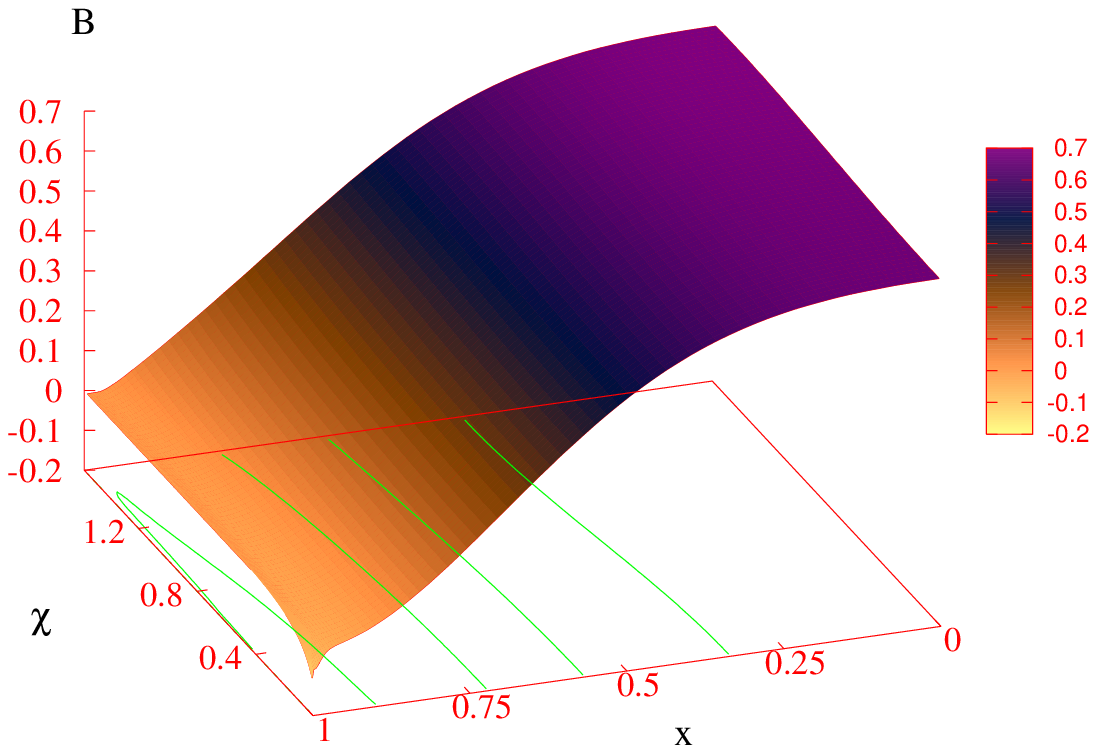,width=8cm}}
\put(-1,12){\epsfig{file=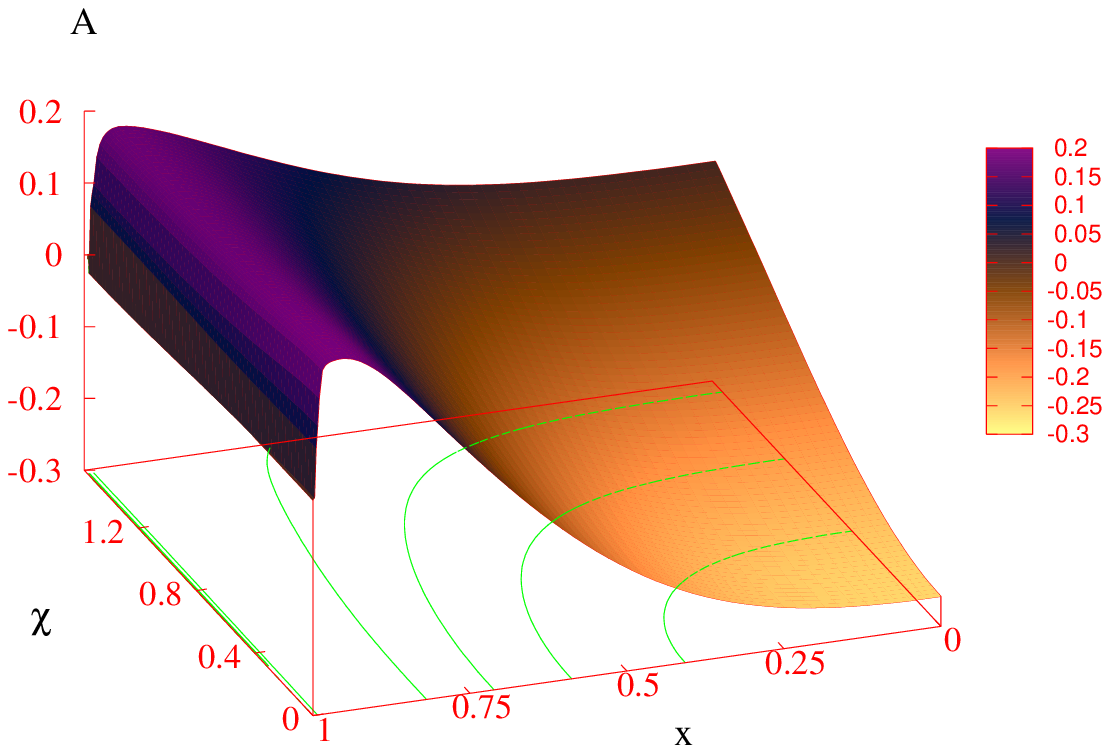,width=8cm}}
\put(7,12){\epsfig{file=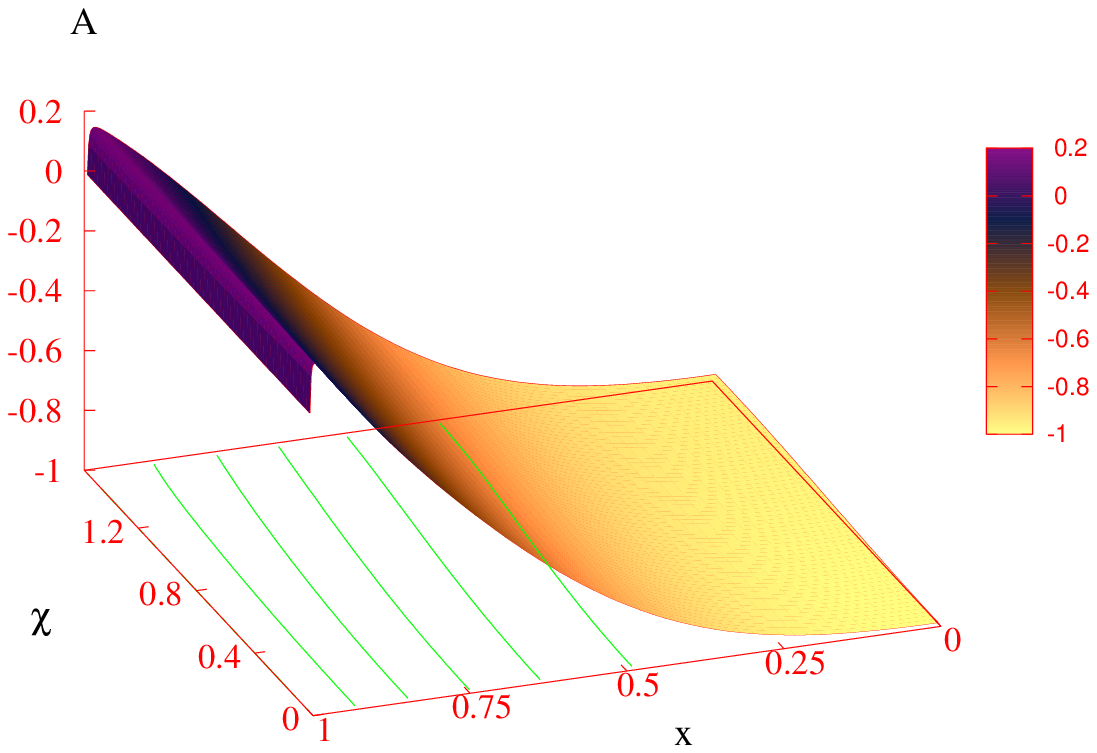,width=8cm}}\end{picture} 
\\
\\
{\small {\bf Figure 1.}
The metric functions 
$A$, $B$, and $C$ 
are shown as functions of the compactified radial coordinate $x=\rho/(1+\rho)$ 
and the angular variable $\chi$
for a vacuum configuration with $L=1$ (left) 
and a charged one with $L=0.24,~q=1.73$ (right).
} 
\end{figure}
In this case, the solver has converged and provided numerical output 
with good accuracy\footnote{The computed relative error of the ``solution''
(truncation error) was on the order of $0.001$ in this case.}.
The functions $A$, $B$ and $C$ have a nontrivial shape, 
but their magnitude is small, almost zero.
Next, we have started a new iteration and employed this data as initial guess
for a slightly larger value of $L$.
Again, the solver has converged and provided apparently reasonable
numerical output, which in turn has been used as input 
for the next iteration for another slightly larger value of $L$, etc.
For all configurations with sufficiently small values of $L$, 
we have found good agreement of the extracted physical properties 
with those reported by Kudoh et al.~\cite{Kudoh:2003xz,Kudoh:2003vg},
obtained within their numerical scheme.

In principle, working in small steps, 
one expects to find solutions for arbitrarily large values of $L$ in this way.
However, for any grid choice employed, 
we have noticed that the numerical accuracy appeared to be progressively 
deteriorating with increasing $L$.
Around $L\simeq 0.3$ finally, the numerical errors 
have turned unacceptably large for the 
configurations\footnote{This has also been manifest
in the constraint equations $G_\chi^\rho$ and $G_\rho^\rho-G_\chi^\chi$, 
implying large errors in the evaluation of a Hawking temperature.},
while for still larger values of $L$ the numerical errors 
have accumulated further, until finally
convergence of the numerical scheme has been lost.

In our scheme,
the numerical problems appear to originate mainly in the asymptotic 
region\footnote{In previous numerical work 
numerical problems were largely associated with the symmetry axis $\chi=0$.}.
As $L$ increases, the metric functions
start to develop an increasingly unnatural shape 
for large values of the radial coordinate (typically for $\rho> 10 \ell$)
and all values of $\chi$ 
(although in some functions this feature is more pronounced 
for $\chi\to 0$). 
This unnatural behaviour manifests itself
in the occurrence of ``oscillations'' 
of the metric functions in the far field\footnote{A similar behaviour 
was noticed by Yoshino \cite{Yoshino:2008rx} (see his Figure 3).}.
The amplitude of these ``oscillations'' increases with increasing $L$. 
At the same time these ``oscillations'' start 
at increasingly smaller values of the radial coordinate.
We illustrate this unnatural behaviour of the metric functions
close to the AdS horizon for a typical $d=5$ calculation
(with large numerical errors)
in Figure 1 (left) and also in Figure 2 (the case with $q=0$).

This unnatural behaviour 
appears for all grid choices 
and all metric parametrizations considered, 
including the one employed in \cite{Kudoh:2003xz,Kudoh:2003vg,Yoshino:2008rx}.
Also, quite interesting, when working with a finite grid 
with cut-off radius  $\rho_{max}\simeq 10 r_0$  
(the typical case considered in \cite{Kudoh:2003xz}),
one usually fails to detect this oscillatory behaviour, 
since it appears only for larger values of the radial coordinate.
(A distinct dependence of the numerical results on the value of $\rho_{max}$ 
was already noticed by Yoshino \cite{Yoshino:2008rx},
representing an essential part of the nonsystematic error
present in his calculations.)
 
Moreover, since the numerical error increases gradually, 
it is impossible to identify a critical value $L_c$,
such that one could claim non-existence of the solutions for $L>L_c$.
Thus the ``oscillations'' are likely to exist for any value of $L$.
We surmise that for very small values of $L$
such a behaviour is located at values of $x$ very close to unity 
($i.e.$ at very large values of $\rho$), 
and that the amplitude of the ``oscillations'' is at the same time very small 
(and thus difficult to detect within the numerical approach employed).

\subsubsection{ $U(1)$ field on the brane}

It is interesting to examine whether the above results hold
also for more general cases.
In particular, one would like to know whether
the unnatural asymptotic behaviour, that we have noticed in the vacuum case,
survives in the presence of matter fields on the brane
or whether it can be circumvented due to their presence.

The simplest case one may think of corresponds to a spherically 
symmetric static Maxwell field living on the brane. 
As one can see from eqs.~(\ref{bc-brane2}), 
although the bulk equations are the same as in the vacuum case, 
the presence of the U(1) field leads to a different set
of boundary conditions on the brane 
and thus to a different geometry in the bulk.
Naively, one may then expect the metric on the brane to correspond 
to a RN black hole with corrections.

Our initial hope has been that the presence of an electric charge 
on the brane would make the numerical scheme more stable.
However, in contrast to our expectations,
our results for typical charged 
configurations have turned out to be qualitatively similar 
to the vacuum case\footnote{We have considered charged configurations
only in $d=5$ dimensions.}.
For a given value of $L$,
we have employed the corresponding vacuum data as the initial guess
for the numerical integration of the equations in the presence
of a small electric charge $Q$.
Subsequently, we have slowly increased the value of $Q$,
and thus the second dimensionless parameter $q$.

For small values of $L$ (typically $L<0.1$), 
the solver has converged and has provided apparently reasonable
numerical output, as in the vacuum case.
In particular, the shape of the functions 
appears to be rather insensitive to the value of $q$, 
while their magnitude is increasing with $q$. 
When extracting the Hawking temperature
and the horizon area from the numerical data
for these small values of $L$,
we have observed that
for fixed $L$ the Hawking temperature decreases with increasing $q$
as expected,
while the horizon area increases for the 
``supposed'' bulk black hole 
as well as for the associated ``supposed'' black hole on the brane.

However, for any value of $q$ employed, 
we have seen the numerical accuracy 
of the calculations to strongly deteriorate with increasing $L$
as in the vacuum case,
until at some stage the numerical solver has stopped to converge.
Moreover, the behaviour of the metric functions for large values of $\rho$
is also qualitatively similar to what we have found for $q=0$,
while the magnitude of the ``oscillations'' 
in the asymptotic region even increases with increasing $q$, 
as seen in Figure 2.
This unnatural behaviour appears to be generic 
for any choice of the nonequidistant grid employed in the integration,
and we conclude that all pathologies observed 
for vacuum configurations are also present in the charged case.
Clearly, this puts the existence of such charged black hole solutions 
into strong doubt.

\begin{figure}[t!]
\setlength{\unitlength}{1cm}
\begin{picture}(15,18)
\put(-1,0){\epsfig{file=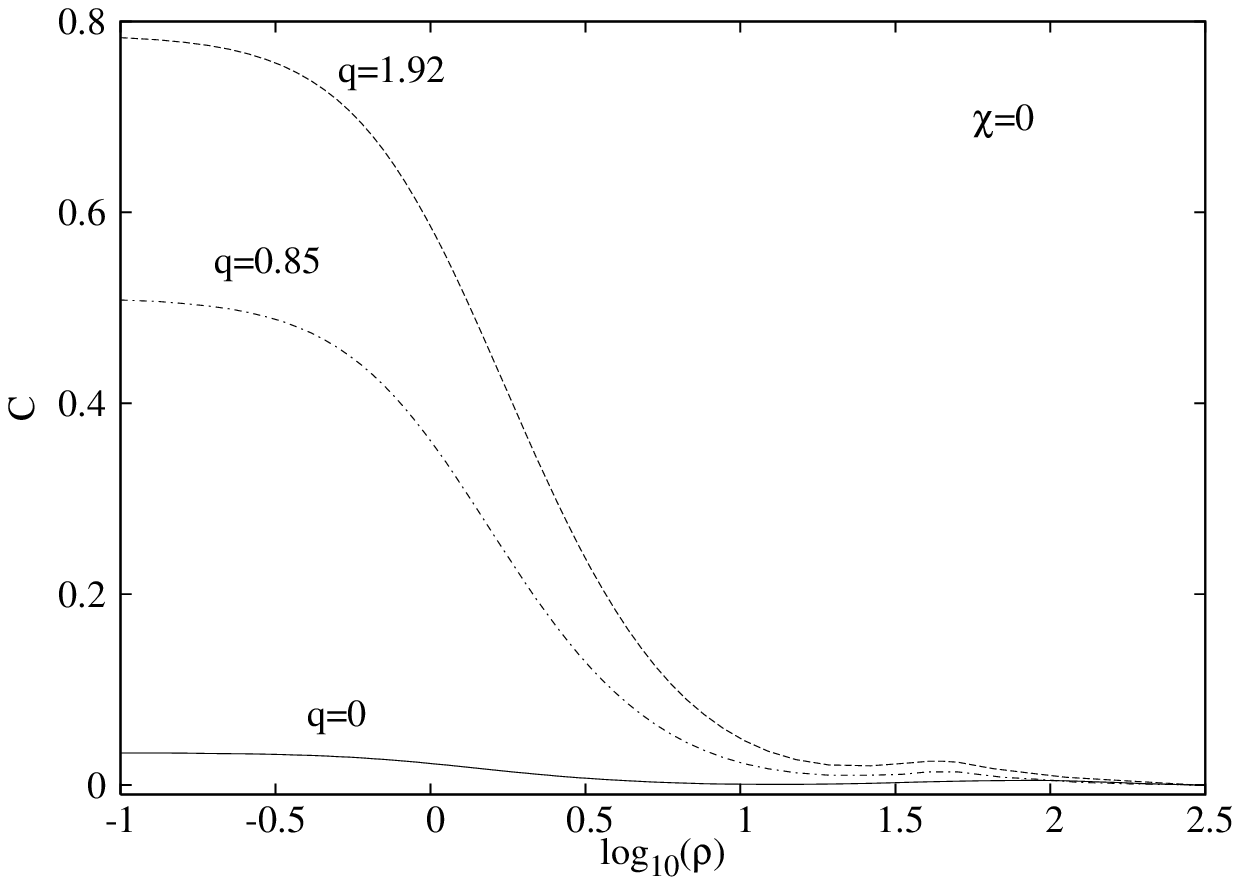,width=8cm}}
\put(7.5,0){\epsfig{file=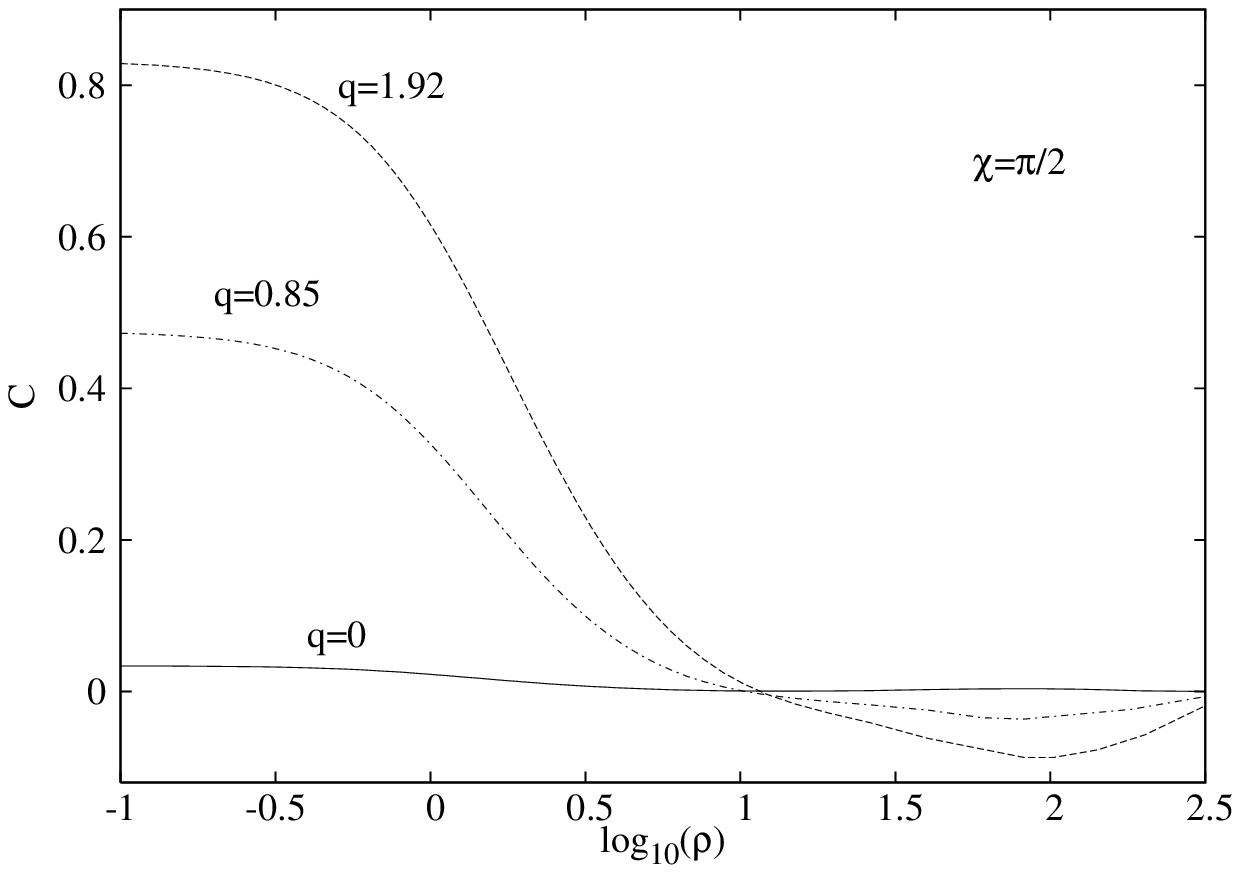,width=8cm}}
\put(-1,6){\epsfig{file=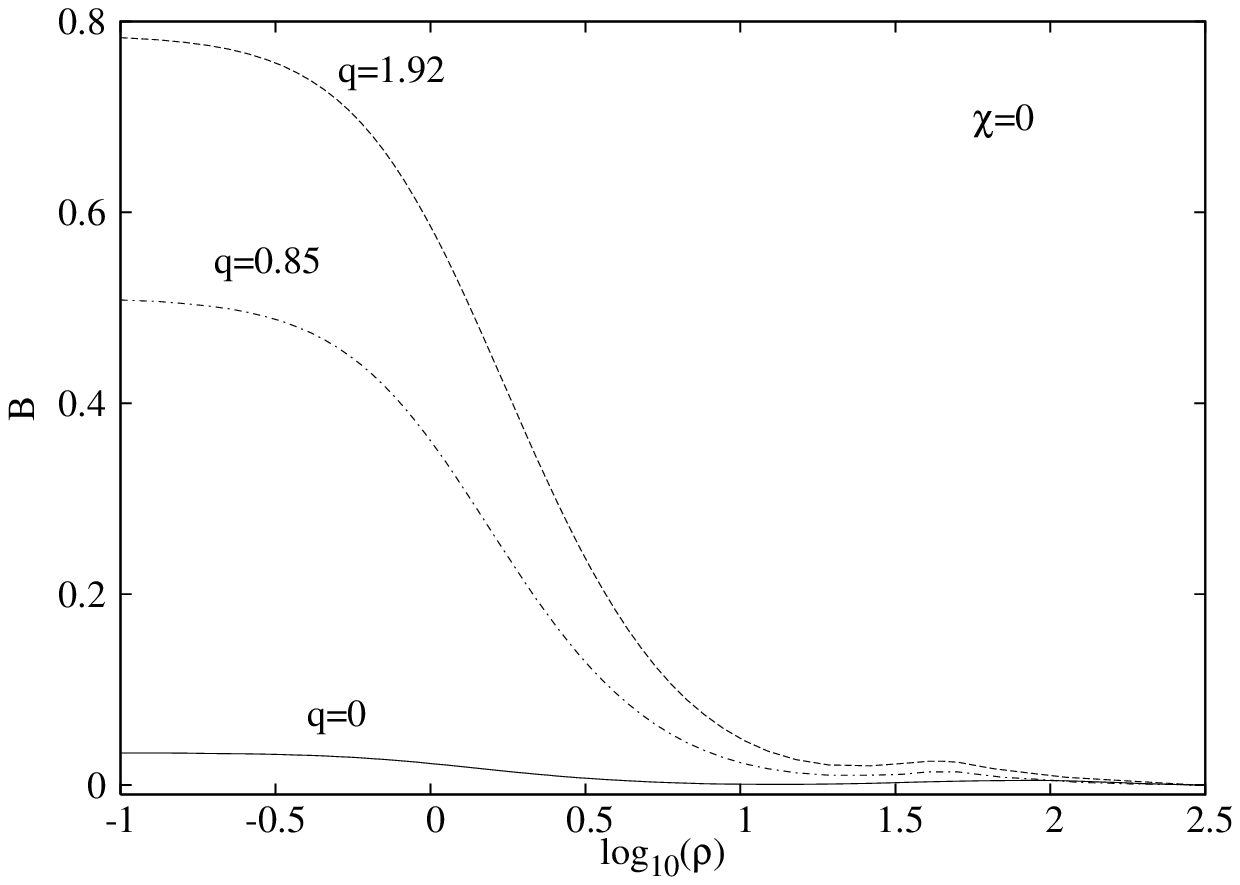,width=8cm}}
\put(7.5,6){\epsfig{file=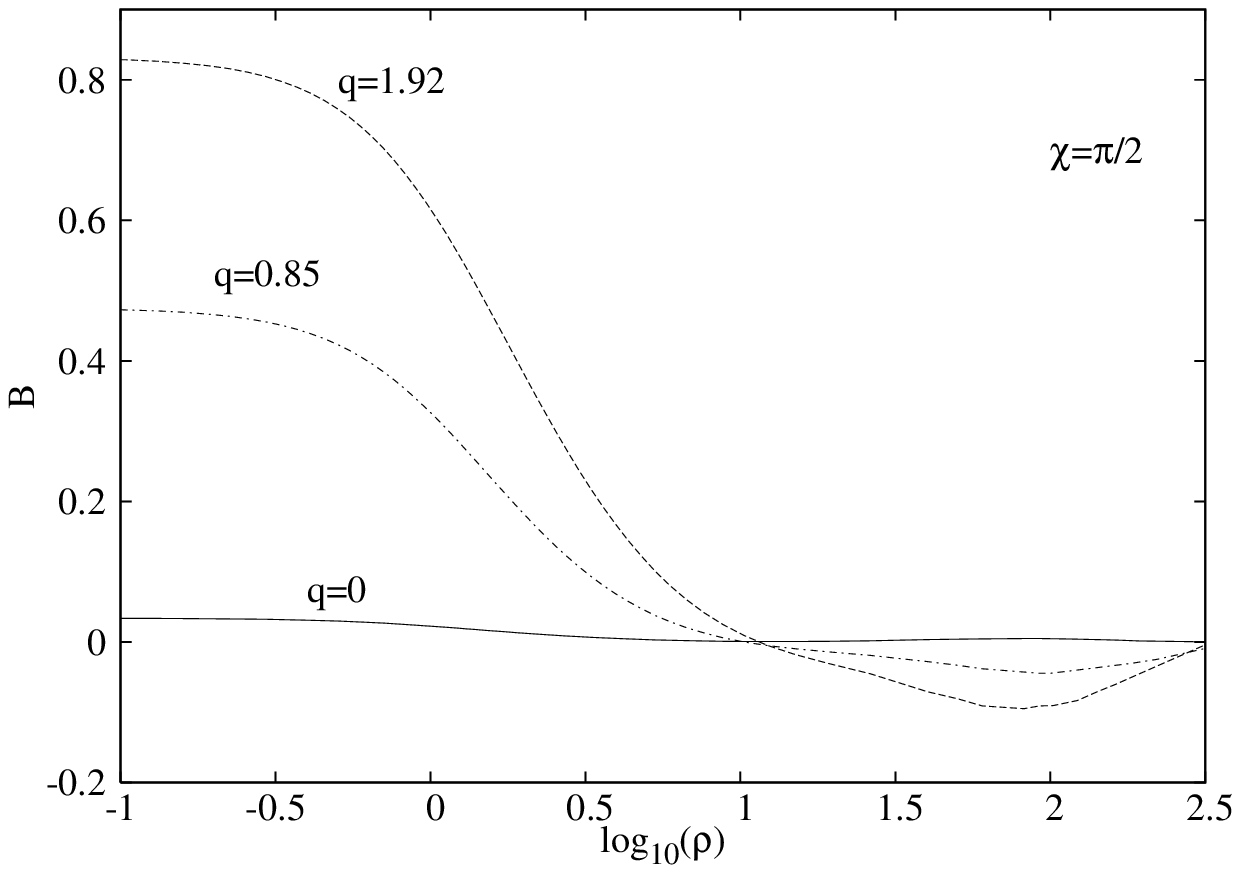,width=8cm}}
\put(-1,12){\epsfig{file=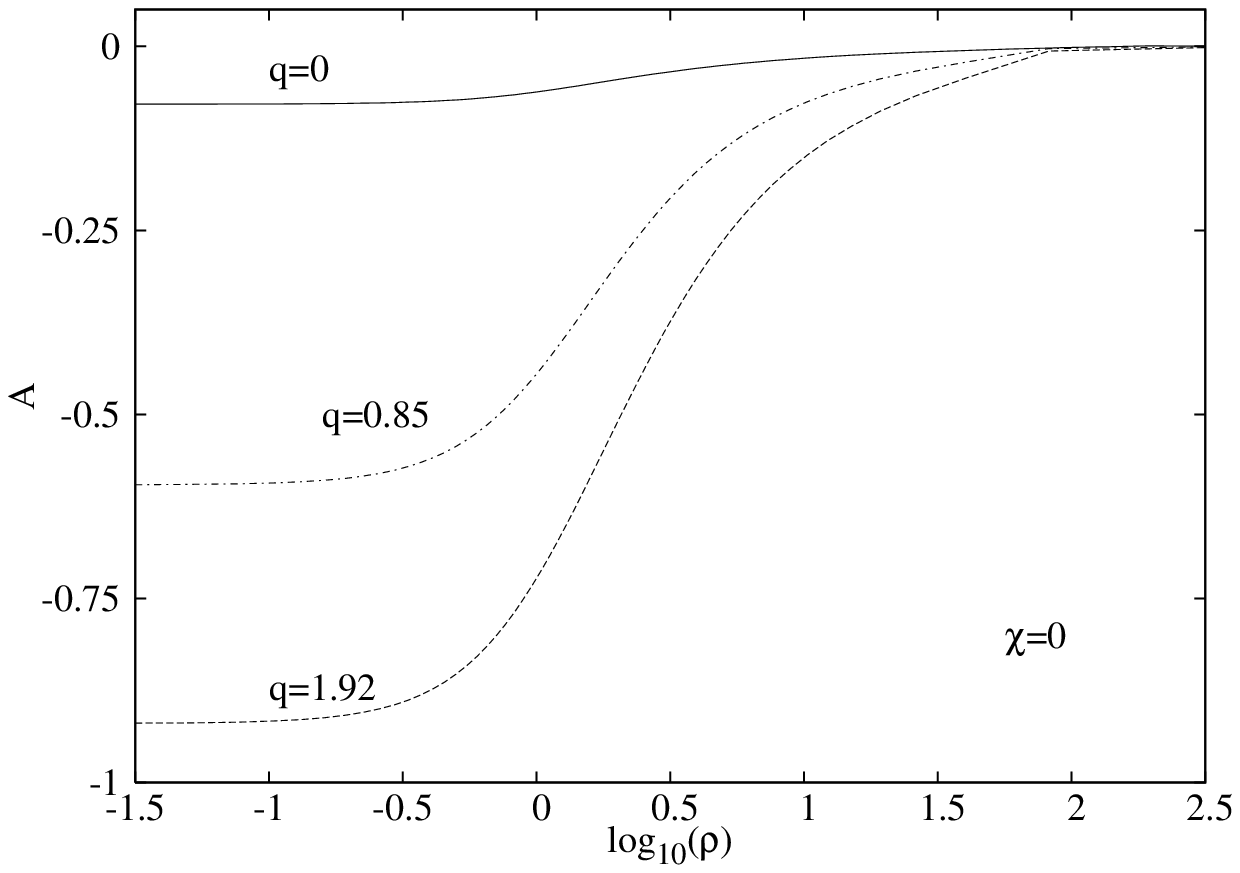,width=8cm}}
\put(7.5,12){\epsfig{file=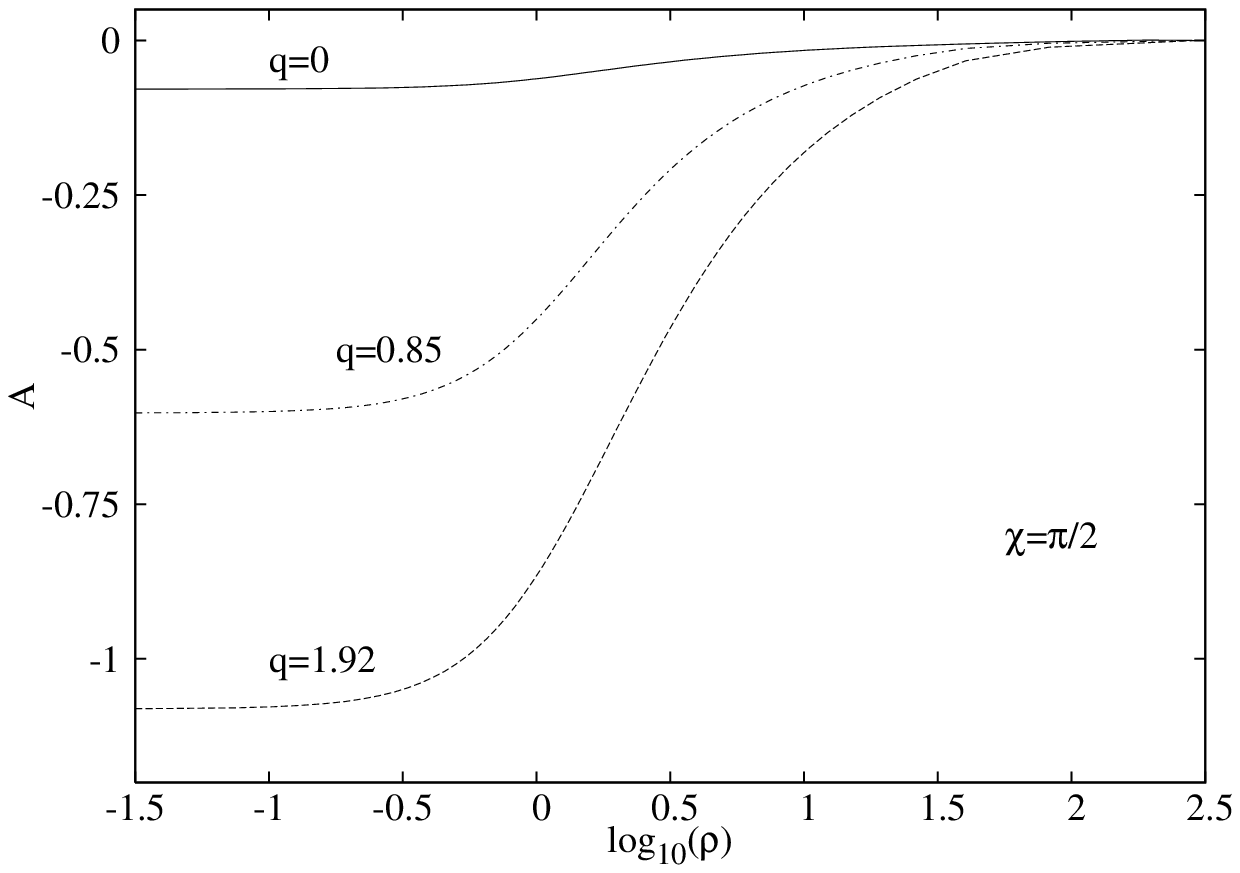,width=8cm}}
\end{picture}
\\
\\
\\
{\small {\bf Figure 2.}
The metric functions 
$A$, $B$ and $C$ are shown at $\chi=0$ (on the symmetry axis)
and at $\chi=\pi/2$ (on the brane) 
for configurations for $L=10$ and three values of the charge parameter $q$.}
\end{figure}

We would like to emphasize that all of the configurations 
obtained numerically yield a nonvanishing Hawking temperature,
and thus would correspond to non-extremal black holes
(if they were indeed solutions).
Evaluated according to eq.~(\ref{TH}),
the expression for the Hawking temperature 
does not yield a strictly constant value,
but typically varies slightly (because of the numerical error),
as long as $L$ and $q$ are sufficiently small.
For the charged configurations obtained here, however,
the numerical error makes $|A(0,\chi)-B(0,\chi)|$
(and thus the temperature) deviate appreciably from a constant value.
In fact, the variations become the larger the larger the charge $q$,
whereas in principle the equations should guarantee a constant 
value of the temperature.
It is thus not surprising to see the numerical integration
fail to converge for some critical value of the charge.

On the other hand, with increasing $q$,
the ``supposed'' Hawking temperature of the configurations decreases,
and configurations with appropriately large values of $q$
would be expected to correspond to near-extremal solutions.
Indeed, this reasoning suggests that, similar to GR, 
extremal solutions might exist,
which possess a maximal value of the electric charge 
for a given event horizon radius, 
and whose temperature would be zero.
However, the study of black holes which are extremal 
or close to extremality is a difficult numerical problem 
already for situations, 
where the existence of solutions is not controversial.
Consequently, the existence of $extremal$ braneworld black holes 
in the RS model is an open problem, which we cannot attack
within our current numerical scheme.
However, we will address the problem from a different direction in the next Section.

\subsubsection{The interpretation of the results}
We think that the interpretation of our results can only be analogous
to the interpretation proposed by Yoshino \cite{Yoshino:2008rx}.
In particular, we would like to consider the following possibilities:

First, one might suspect that our approach and/or the numerical methods are wrong.
However, similar methods\footnote{In 
particular, the treatment of the behaviour of the metric functions on the 
axis $\chi=0$ and at infinity has been similar to that used in this work.} 
were used in the past to solve a variety of problems, 
see $e.g.$~\cite{Kleihaus:2006ee,Kleihaus:1997ic,Radu:2008pp}. 
Moreover, we have extensively tested these numerical routines
to recover numerous exact solutions in GR and field theory.
At the same time, some of the new solutions derived 
by using the code FIDISOL/CADSOL were rederived subsequently 
by other groups with different numerical methods.
Therefore we think that this hypothesis can be safely excluded.

The second possibility would be that static solutions exist, 
but that they are very hard to find, forming isolated
``islands'' in the parameter space.
Moreover, these ``islands'' would be disconnected 
from the $L=0$ Schwarzschild black hole used as initial guess 
in the numerical attempts to construct such solutions\footnote{Such 
examples are known for numerical solutions in field theory, 
the case of vortons (which are $d=4$ flat space toroidal solitons 
somewhat analogous to $d=5$ black rings) being perhaps the most notorious. 
There  several disconnected branches of numerical solutions are known to exist,
which do not possess a static limit \cite{Radu:2008pp,Battye:2008mm}.}.
In this setup, the numerical results found would be numerical artifacts, 
since the true solutions would require a starting profile 
different from the Schwarzschild black hole.
However, we consider this possibility as counter-intuitive and unlikely.

In our opinion, the most likely scenario is that 
{\it there are no static solutions} of the 
bulk Einstein equations (\ref{neq-A})-(\ref{neq-C})
with the boundary conditions (\ref{bc1})-(\ref{bc3}) and (\ref{bc-brane2}).
Moreover, in agreement with the conjecture put forward 
by Emparan et al.~\cite{Emparan:2002px} and Tanaka \cite{Tanaka:2002rb},
all nonextremal black holes on the brane would be dynamical objects.
If this were indeed the case, 
then the numerical results could be interpreted as follows:
what we have found should correspond to static approximations 
of some yet unknown time-dependent configurations.
The full solutions for black holes on the brane 
would be dynamical, quantum corrected, evaporating black holes; 
thus also the bulk solutions would depend on time.
The systematic occurrence of the observed unnatural behaviour 
of the metric functions close to the AdS horizon would be the result 
of an inappropriate static metric ansatz.
For black hole with sizes small compared to the AdS length scale, 
the pathological behaviour at infinity
would not be seen due to the lack of numerical accuracy.
In other words, for small values of $|\Lambda|$, 
the hyperbolic character of the equations would not become manifest,
and the solver would fail to see the quantum corrections.  
However, as the value of $L=r_0/\ell $ increases, 
the neglected dynamical terms would destabilize the 
numerics\footnote{We note
that the observed oscillatory behaviour of the configurations
is not predicted by perturbation theory 
(see \cite{Kudoh:2003xz} and references therein).
However, we suspect that this is a deficiency of the perturbation theory.
A somewhat similar situation is encountered in Einstein gravity 
coupled to nonabelian matter fields. 
There perturbation theory predicted the existence of $d=4$ 
rotating Einstein-Yang-Mills solitons \cite{Volkov:1997qb}.
Such configurations were, however, ruled out by nonperturbative arguments. 
}.
This interpretation should hold for both vacuum and charged configurations.

\section{Extremal solutions: the near-horizon geometry}

\subsection{The problem}

The discussions in the previous Section concern the case of nonextremal
static black holes possessing a {\it nonzero} Hawking temperature.
However, these findings cannot be used to argue against the existence of 
static {\it extremal} configurations.
In contrast, static {\it extremal} configurations
would have zero temperature 
and therefore would not Hawking radiate. 
Thus the arguments put forward by 
Emparan et al.~\cite{Emparan:2002px} and Tanaka \cite{Tanaka:2002rb}
against the existence of static black holes on the brane 
would be circumvented.
From this point of view,
arbitrarily large {\it extremal} black holes 
might indeed exist on the brane.

In principle, the metric ansatz (\ref{metric-ansatz}) 
could still be used in an attempt to 
numerically construct extremal black holes, 
after replacing the background function $F$ 
by $F(r)=(1-(r_0/r)^{d-3})^2$.
(Then the bulk metric would still depend on the coordinates $r,\chi$.) 
However, the numerical construction of extremal black holes 
is a highly non-trivial task\footnote{In
fact, we are not aware of any successful numerical construction 
of extremal black holes, obtained as solutions 
of second order partial differential equations.}.
Therefore we do not attempt here to construct such bulk configurations
numerically.
Instead, in the following we only address a simplified problem
and concentrate on the near-horizon region of ``potential'' extremal black holes. 

Such a restricted study may well provide hints
on the properties of the full black hole solutions
without solving for the full metric\footnote{Such an
approach has been used recently to investigate properties 
of higher dimensional extremal black holes 
with a nonspherical horizon topology, see $e.g.$~\cite{Figueras:2008qh}.}.  
However, we would like to emphasize
that this study could only allow us to rule out 
possible full black hole solutions, 
while it cannot prove their existence,
since for a given near-horizon geometry 
there need not be a corresponding full black hole solution.
 
Let us start by recalling that in GR
an extremal RN solution in $D$-dimensions 
(where later $D=d-1$)
has a near-horizon geometry $AdS_2\times S^{D-2}$, 
which is also a solution of the equations of motion.
For a Lagrangian density $L=R-F^2$, 
the corresponding near-horizon line element reads 
\begin{eqnarray}
\label{GR1}
ds^2=L_{1(GR)}^{2} d\Sigma^2_2+L_{2(GR)}^{2} d\Omega_{D-2}^2,
\end{eqnarray}
where
\begin{eqnarray}
\label{GR2}
L_{1(GR)}^{2} =\frac{2(D-3)}{D-2}Q^2,~~~L_{2(GR)}^{2}=\frac{2(D-3)^3}{D-2}Q^2,
\end{eqnarray}
$d\Sigma^2_2=dx^2/x^2-x^2 dt^2$ is the line element 
of the two-dimensional AdS space,
and the gauge field is $F_{x t}=Q$.
(Note, that the electric charge ${\mathcal Q}$ 
of the bulk RN black holes is 
${\mathcal Q}=Q L_{2(GR)}^{D-2}/L_{1(GR)}^{2}$ (up to a volume factor).) 
The entropy of this solution is
\begin{eqnarray}
\label{S-brane}
S_{D(GR)}=\frac{V_D}{4 G_D} L_{2(GR)}^{D-2}.
\end{eqnarray}

Turning now to black holes on the brane,
we expect that the near-horizon geometry 
of static extremal black holes on the brane in the RS model 
will - analogously - have a near-horizon geometry 
$AdS_2 \times S^{d-3}$ 
(where $AdS_2$ and $S^{d-3}$ have constant radii), 
which should then be the induced metric on the brane.
Thus we assume that this metric has a form like eq.~(\ref{GR1}), 
with, in general, different values for the constant coefficients, 
but recovering eqs.~(\ref{GR2}) and (\ref{S-brane}) 
in a certain limit.

These considerations correspond in fact to the approach 
of Kaus and Reall \cite{Kaus:2009cg} 
used to study $d=5$ extremal black holes,
charged with respect to a Maxwell field on the brane.
Kaus and Reall \cite{Kaus:2009cg}
found that the GR results are recovered for large black holes.
(We further note, that Suzuki et al.~\cite{Suzuki:2010kv} examined the case of
extremal black holes in a braneworld with a cosmological constant.)

\subsection{The bulk: equations and asymptotics}
The considerations above lead us to propose the following line element
for the near-horizon limit of an extremal bulk black hole, 
\footnote{We 
employed a similar metric ansatz and the same numerical methods
in our previous studies \cite{Mann:2006yi,Kleihaus:2010am} 
of asymptotically AdS solitons with a nonstandard asymptotic structure.}
\begin{eqnarray}
\label{n-b-m}
ds^2=\frac{d\xi^2}{f(\xi)}+a(\xi) d\Sigma_2^2+\xi^2 d\Omega_{d-3}^2,
 \end{eqnarray}
containing two metric functions $a(\xi)$ and $f(\xi)$.
Here the coordinate $\xi\geq 0$
is the coordinate normal to the brane, 
and the brane is located at some $\xi_0>0$. 
Thus the coordinate $\xi$ 
is proportional to $\sin \chi$
in the bulk parametrization eq.~(\ref{metric-ansatz})\footnote{The
coordinate $r$ in eq.~(\ref{metric-ansatz})
becomes the coordinate $x$ in the $AdS_2$ parametrization
$d\Sigma^2_2=dx^2/x^2-x^2 dt^2$
after taking the near-horizon limit.}.

The functions $a(\xi)$ and $f(\xi)$ are solutions of the differential equations
\begin{eqnarray}
\label{eqs-bulk}
&&a''
-\frac{a'^2}{a^2}
+\frac{(d-5)(d-4)}{2 \xi^2}(1-\frac{1}{f})
+\frac{(d-4)af'}{2\xi f}
+\frac{(d-4)a'}{\xi}
+\frac{a'f'}{2f}
+\frac{1}{f}
+\frac{\Lambda a}{f}=0,
\\
\nonumber
&&f'+\frac{4\Lambda \xi}{d-2}+\frac{2(d-4)}{\xi}(f-1)+\frac{2fa'}{a}=0,
 \end{eqnarray}
together with the constraint equation
 \begin{eqnarray}
\label{constr}
a'^2
+\frac{2(d-3)(d-4)a^2}{\xi^2}(1-\frac{1}{f})
+\frac{4a(1+\Lambda a)}{f}
+\frac{4(d-3)a a'}{\xi}=0.
\end{eqnarray}
These equations have the following exact solutions,
\begin{eqnarray}
\label{sol1}
 f(\xi)=1+\frac{\xi^2}{\ell^2},~~ a(\xi)=\xi^2+\ell^2,
\end{eqnarray}
which corresponds to $AdS_d$ in coordinates adapted to a foliation 
by $AdS_2 \times S^{d-2}$ hypersurfaces,
and
\begin{eqnarray}
\label{sol2}
 f(\xi)=1+\frac{d-1}{d-3}\frac{\xi^2}{\ell^2},~~ a(\xi)=\frac{\ell^2}{d-1},
\end{eqnarray}
which corresponds to $AdS_2 \times H^{d-2}$.

For the general solutions of these equations smoothness at $\xi=0$
requires that $a(\xi)$ and $f(\xi)$
possess a Taylor series expansion there, consisting of even powers of $\xi$ only, 
with $a(0)>0$ and $f(0)=1$. 
To order $\xi^4$, the small $\xi$ expansion reads
\begin{eqnarray}
\label{initial}
&&
a(\xi)=a_0-\frac{d-2+2a_0 \Lambda}{(d-2)^2 }\xi^2
+\frac{(d-2+2a_0 \Lambda)((d-1)(d-2)+2a_0\Lambda)}{a_0(d-2)^4(d-3)d}\xi^4+\dots,
\\
\nonumber
&&
f(\xi)=1+\frac{2(d-2-a_0(d-4) \Lambda)}{a_0(d-2)^2(d-3) }\xi^2
+\frac{2(d-2+2a_0 \Lambda)((d-1)(d-2)+2a_0\Lambda)}{a_0^2(d-2)^4(d-3)d}\xi^4+\dots~.
\end{eqnarray}

One can also construct an approximate solution for large values of the
ratio $\xi/\ell$ 
(although we shall see that configurations with this asymptotics 
exist for a restricted set of initial data, $a(0)>\ell^2/(d-1)$ only), 
which is useful in our analytical study of large black holes.
For even $d$, this solution reads
\begin{eqnarray} 
\label{even-inf}
a(\xi)&=&U \ell^2
\left (
\frac{\xi^2}{\ell^2}+\sum_{k=0}^{(d-4)/2}a_k(\frac{\ell}{\xi})^{2k}
-M(\frac{\ell}{\xi})^{d-3}
\right)+O(1/\xi^{d-2}),
\\
\nonumber
f(\xi)&=&\frac{\xi^2}{\ell^2}+\sum_{k=0}^{(d-4)/2}f_k(\frac{\ell}{\xi})^{2k}
-2M(\frac{\ell}{\xi})^{d-3}+O(1/\xi^{d-2}),
\end{eqnarray}   
where the parameter $U$ corresponds to the asymptotic ratio 
of the radius of $AdS_2$ to that of $S^{d-3}$. 
Also, $a_k$ and $f_k$ are constants depending on  
the spacetime dimension $d$ and $U$ only.   
Specifically, one finds
\begin{eqnarray}
\label{inf2}  
&&a_0=\frac{1+(d-4)U}{U(d-3)},~~
a_1=-\frac{(d-4)(1-U)(1+(d-4)U)}{(d-2)(d-3)^2(d-5)U^2},
\\
\nonumber
&&a_2=\frac{(d-4)(1-U)(1+(d-4)U)\left(2(d-5)d-4 
+(d-4)(26+(3d-23)d)U \right)}
{3(d-2)^2(d-3)^3(d-5)(d-7)U^3}~,
\end{eqnarray} 
and
\begin{eqnarray}
\label{inf3}  
&&f_0=\frac{2 +(d-1)(d-4)U}{U(d-2)(d-3)},~~
f_1=-\frac{2(d-4)(1-U)(1+(d-4)U)}{(d-2)(d-3)^2(d-5)U^2},
\\
\nonumber
&&f_2=\frac{2(d-4)(1-U)(1+(d-4)U)\left( d(d-7)+8
+(d-4)(11+(d-8)d)U \right)}
{(d-2)^2(d-3)^3(d-5)(d-7)U^3}~,
\end{eqnarray}  
their expressions becoming more complicated for higher order $k$,  
without exhibiting a general pattern.
The corresponding expansion for odd values of the spacetime 
dimension is more complicated, 
with $\log (\xi/\ell)$ terms in the asymptotic expression
\begin{eqnarray}
\label{odd-inf}
a(\xi)&=&U
\left (
\frac{\xi^2}{\ell^2}+\sum_{k=0}^{(d-5)/2}a_k(\frac{\ell}{\xi})^{2k}
+q\log(\frac {\ell}{\xi}) (\frac{\ell}{\xi})^{d-3}
+(-M+\beta)(\frac{\ell}{\xi})^{d-3}
\right)
+O(\frac{\log \xi}{\xi^{d-1}}),
\\
\nonumber 
f(\xi)&=&\frac{\xi^2}{\ell^2}+\sum_{k=0}^{(d-5)/2}f_k(\frac{\ell}{\xi})^{2k}
+2q \log (\frac {\ell}{\xi}) (\frac{\ell}{\xi})^{d-3}
-2M(\frac{\ell}{\xi})^{d-3}+O(\frac{\log \xi}{\xi^{d-1}}),
\end{eqnarray}   
where $a_k$ and $f_k$ are still given by eqs.~(\ref{inf2}) and (\ref{inf3}), 
while $q$ and $\beta$ are two new constants depending on $\ell$ and $d$,
that can be expressed in a compact form as
\begin{eqnarray}
\label{inf4}
&&
\beta=0\delta_{5,d}
-\frac{3(1-U)^2(1+3U)}{3200 U^3}\delta_{7,d}
+\frac{25 (1-U)^2(1+5U)(157  +725U)}{42674688}\delta_{9,d}+\dots \ ,
\\
\nonumber
&&
q=\frac{1-U^2}{12U^2}\delta_{5,d}
-\frac{3(1-U)(1+3U)(2 +3U)}{800 U^3}\delta_{7,d}
+\frac{5 (1-U)(1+5U)(171 +1100  U+1375 U^2)}{1778112 U^4}\delta_{9,d}+\dots \ .
 \end{eqnarray} 
The parameter $M$ in eqs.~(\ref{even-inf}) and (\ref{odd-inf}) 
is a constant, that can be fixed by numerical calcuations, 
but whose value is not of interest in the present context.
 
We remark, however, that these bulk solutions are interesting in themselves,
since they provide gravitational duals for some CFTs 
in a fixed $AdS_2\times S^{d-2}$ background given by
\begin{eqnarray}
ds^2=\ell^2(U^2 d\Sigma_2^2+d\Omega_{d-3}^2).
\end{eqnarray}  
These configurations have a well defined mass and action 
which can be computed by using the boundary counterterm prescription    
of Balasubramanian and Kraus \cite{Balasubramanian:1999re}.
(In fact, both the mass and action are essentially fixed by the parameter $M$ 
in the asymptotic expansion, eqs.~(\ref{even-inf}) and (\ref{odd-inf}).)
The stress tensor of the dual CFT can also be calculated by
the method of Myers \cite{Myers:1999ps}.
A detailed study of these aspects will be presented elsewhere 
in a more general context.

\subsection{The brane}

We now assume that the brane is located at $\xi=\xi_0$, and, 
following the RS construction,
we keep the region $0 \leq \xi \leq \xi_0$ of the bulk.
The induced metric on the brane is a product of $AdS_2$ and $S^{d-2}$, with
\begin{eqnarray}
\label{f1}
d\sigma^2=L_1^2 d\Sigma_2^2+L_2^2 d\Omega_{d-3}^2,
\end{eqnarray} 
where
\begin{eqnarray}
\label{f1a}
 L_1 =\sqrt{a(\xi_0)},~~~L_2=\xi_0.
\end{eqnarray} 
Similar to the case of Einstein-Maxwell gravity discussed above,
we shall take a purely electric field with $F_{x t}=Q$. 
(Note that, similar to the GR case,
the electric charge of the full solution on the brane would be only proportional to $Q$.)

The Israel junction conditions on the brane, eq.~(\ref{eqs-israeljc}),
yield the relation 
\begin{eqnarray}
\label{cond1}
\frac{a'(\xi_0)}{a(\xi_0)}=\frac{2}{3}
\left(
\frac{2(d-2)}{\ell \sqrt{f(\xi_0)}}-\frac{2d-7}{\xi_0} 
\right),
\end{eqnarray}
together with
\begin{eqnarray}
\label{cond2}
Q^2=\frac{1}{3}(d-2)(d-3)\frac{a^2(\xi_0)}{\ell \xi_0}
\left (
 \sqrt{f(\xi_0)}-\frac{\xi_0}{\ell}
 \right).
\end{eqnarray}
Then the constraint equation (\ref{constr})
implies that
\begin{eqnarray}
\label{const2}
\frac{1}{L_1^2}-\frac{(d-3)(d-4)}{2L_2^2}=\frac{Q^2}{L_1^4}
\left(
\frac{d-5}{d-3}+
\frac{7d-23}{2(d-2)(d-3)^2}\frac{\ell^2Q^2}{L_1^4}
\right),
\end{eqnarray}
(which has been used to test the accuracy of the numerical results).
Moreover, this relation implies $L_2>L_1$, 
thus the $S^{d-3}$ radius is always greater than the $AdS_2$ radius.
 
The entropy of the bulk solution is taken to be one quarter 
of the event horizon area
(note the factor of two which originates from the $Z_2$
symmetry of the problem)
\begin{eqnarray}
\label{S-bulk}
S_d=2\frac{V_{d-3}}{4 G_d}\int_{0}^{\xi_0}d\xi\frac{ \xi^{d-3}}{\sqrt{f(\xi)}}.
\end{eqnarray}

\subsection{The solutions}
\subsubsection{General results}

We have solved the bulk equations (\ref{eqs-bulk}) by imposing the initial
conditions (\ref{initial}) together with the boundary conditions 
on the brane (\ref{cond1}), (\ref{cond2}) for all dimensions between five and nine.
%
\begin{figure}[ht]
\hbox to\linewidth{\hss%
	\resizebox{8cm}{6cm}{\includegraphics{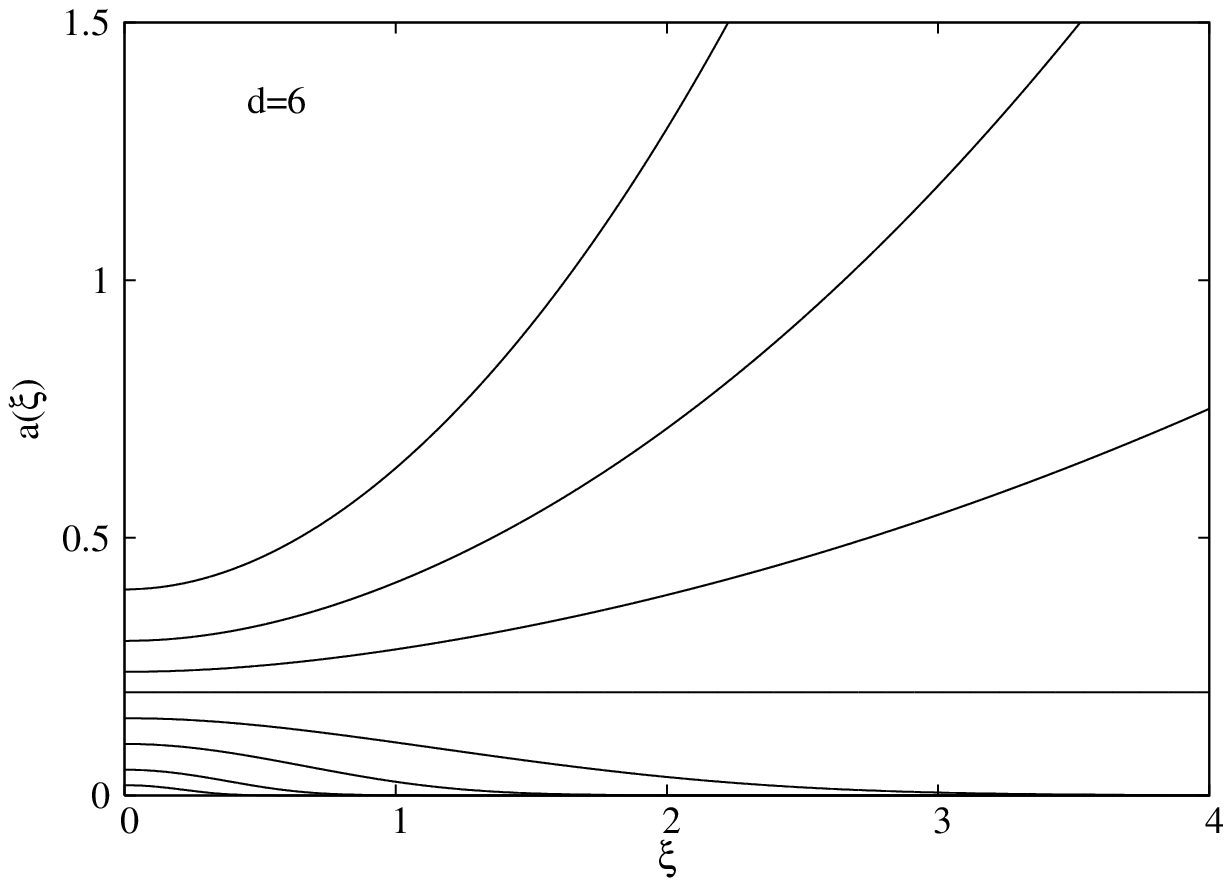}}
\hspace{5mm}%
        \resizebox{8cm}{6cm}{\includegraphics{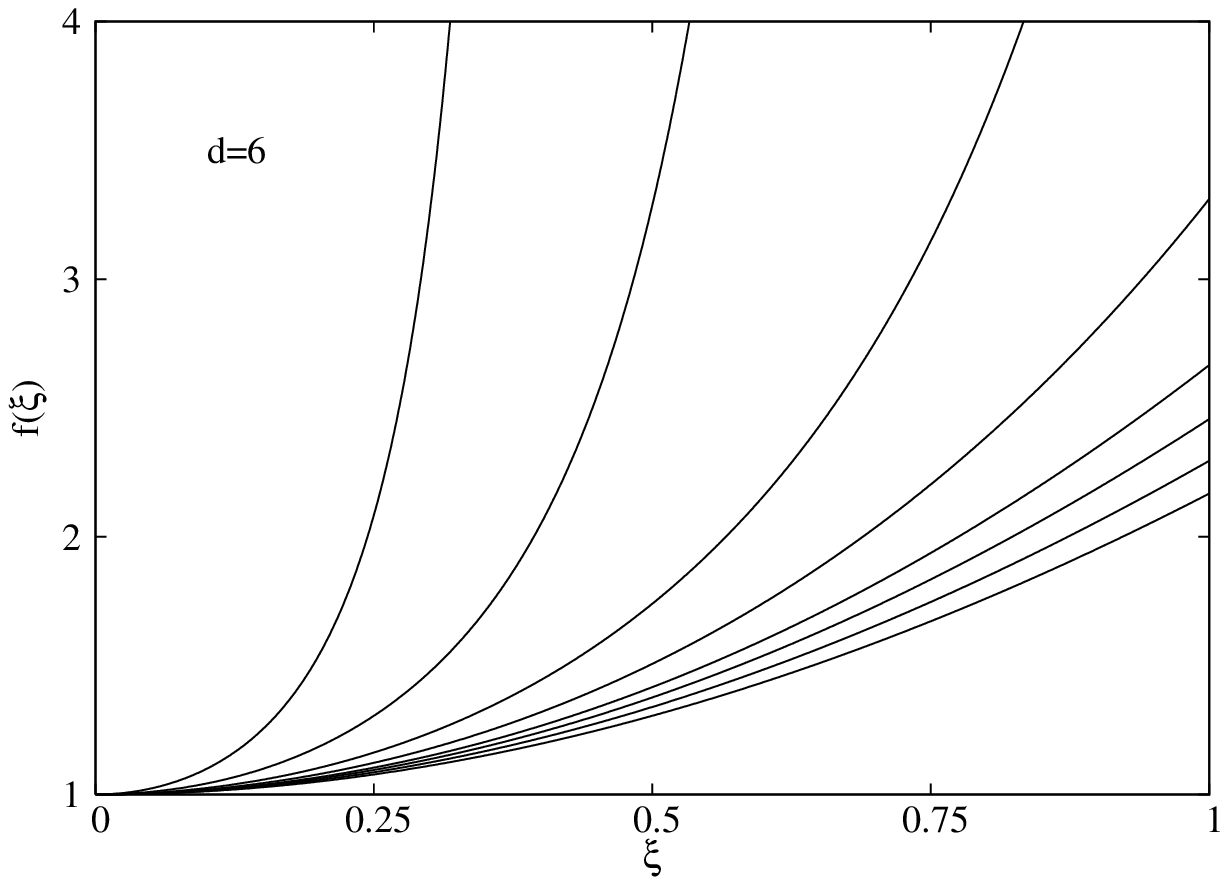}}	
\hss}
\label{diagram3} 
\vspace{0.7cm}
 {\small 
 {\bf Figure 3.}
The functions $a(\xi)$ and $f(\xi)$ are shown for $d=6$ and $\ell=1$.
The initial parameter assumes the values
$a(0)=0.4, 0.3, 0.24, 0.2, 0.15, 0.1, 0.05$ and $0.02$ 
from top to bottom (left figure) and from left to right (right figure).
} 
\end{figure}
\\
Thus such solutions are likely to exist for any higher dimension $d$.
Different from the strategy employed in Section 2,
when trying to solve for the full configurations,
we have here fixed the AdS length scale $\ell=1$
and varied instead the position of the brane, 
$i.e.$, the size of the black hole as given by the parameter $L_2$,
eq.~(\ref{f1a}).

The near-horizon region of the extremal black holes
has then been constructed in several steps as follows.
First, we have solved numerically the Einstein equations (\ref{eqs-bulk}) 
by employing a standard ordinary differential equation solver. 
In particular, for a given value $a_0$, we have evaluated the initial conditions
(\ref{initial}) at $\xi = 10^{-6}$ 
and allowed for a global tolerance of $10^{-12}$, 
when integrating towards large values of $\xi$.
(We note, that we have not encountered any problems in the numerical integration 
of the configurations in this Section.)
Given such a solution of the bulk Einstein equations, 
in the second step we have used the junction condition (\ref{cond1}) 
to evaluate the position $\xi_0$ of the brane.
In the final step, the corresponding value of the charge parameter $Q$
has been obtained from condition (\ref{cond2}).

Our results indicate the existence of a single parameter 
family of solutions of the full problem ($i.e.$, bulk plus brane),
conveniently labelled by the value $a(0)$.
The basic features of these solutions are independent of the dimension $d$.
First, for $0<a(0)\leq \ell^2/(d-1)$, the bulk solutions do not
approach the asymptotic form (\ref{even-inf}), (\ref{odd-inf}).
Instead, the function $a(\xi)$ decreases monotonically 
and vanishes at some finite radius $\xi_{max}$,
which is a curvature singularity, as seen $e.g.$ 
by evaluating the Kretschmann scalar.
However, such configurations are also relevant in the present context, 
since the junction conditions (\ref{cond1}) 
possess a solution with $\xi_0<\xi_{max}$.
(Thus the singularity is outside the physical manifold.)

Solutions of the Einstein equations which for sufficiently large values
of $\xi$ approach the asymptotic form (\ref{even-inf}) and (\ref{odd-inf})
are present for $a(0)>\ell^2/(d-1)$.
In this case, there occurs a maximal value of $a(0)$ as well,
since the junction condition (\ref{cond1}) can be satisfied only 
for $a(0)<a_c$,
where the critical value $a_c$ depends on the dimension.
While $a_c=1$ for $d=5$,
approximate values for $a_c$ are
$0.43$, $0.28$ and $0.21$ for $d=6$, $7$ and $8$, respectively.
Also, 
the position of the brane as given by $\xi_0$ 
is a monotonic function of $a(0)$,
with both $\xi_0/\ell$ and $Q$ diverging as $a(0) \to a_c$,
while the parameter $U$ stays finite in this limit.

The functions $a(\xi)$ and $f(\xi)$ of $d=6$ bulk solutions 
are exhibited in Figure 3
for several values of the initial parameter $a(0)$.
Note that the picture here is generic 
(as seen $e.g.$~by comparing with the figures 
obtained by Kaus and Reall \cite{Kaus:2009cg} for $d=5$).
The exact solution (\ref{sol2}) 
clearly separates the two different types of configurations.

In Figure 4 we exhibit the parameters $L_1$, $L_2$ and $S_d$
for $d=6$ and $7$ solutions,
normalized with respect to the corresponding Einstein-Maxwell results,
eqs.~(\ref{GR2}) and (\ref{S-brane}) with $D=d-1$.  
For any $d$, the parameters $L_1$, $L_2$ and $S_d$ vanish as $Q\to 0$.
 \begin{figure}[ht]
\hbox to\linewidth{\hss%
        \resizebox{8cm}{6cm}{\includegraphics{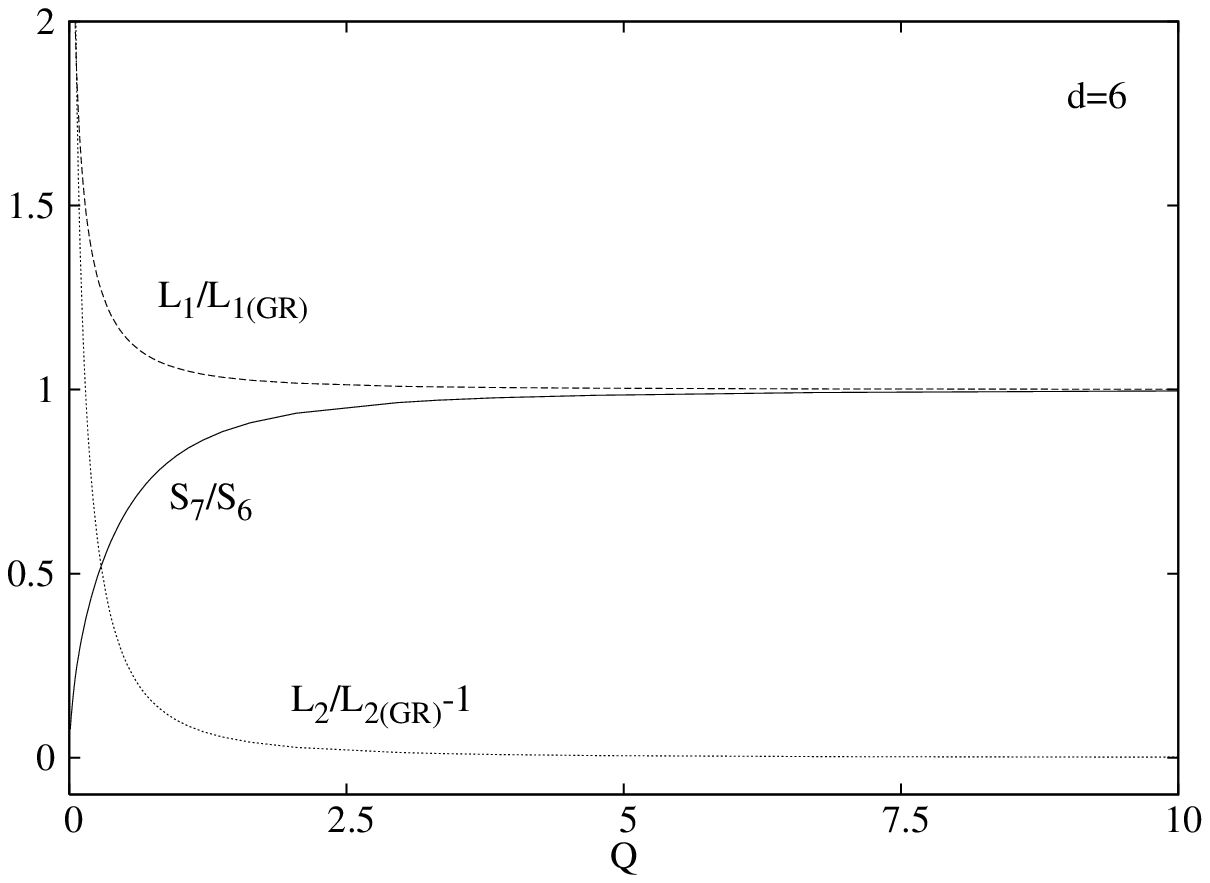}}
\hspace{5mm}%
        \resizebox{8cm}{6cm}{\includegraphics{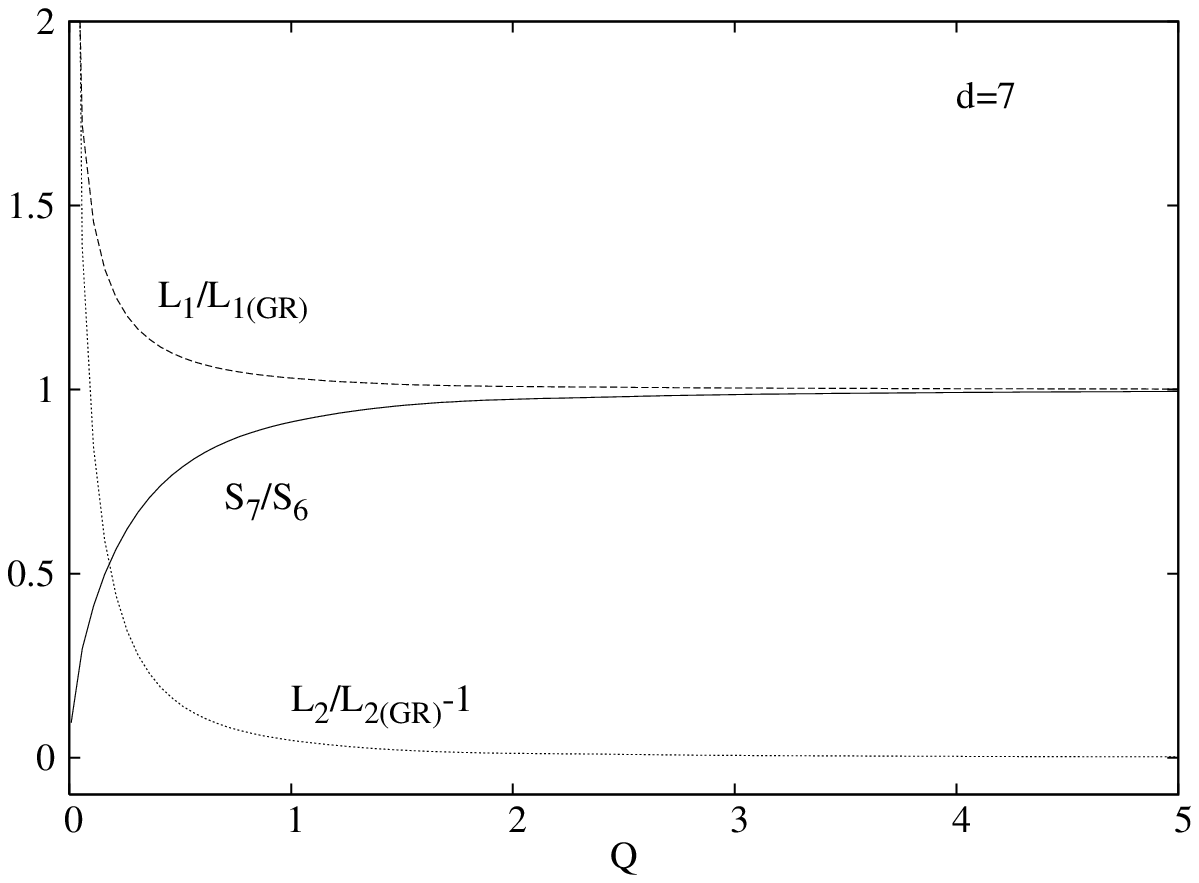}}
\hss}
 \label{diagram4}
\vspace{0.7cm}
 {\small
 {\bf Figure 4.}
The ratios $S_{d}/S_{d-1}$, $L_1/L_{1(GR)}$ and $L_2/L_{2(GR)}$
(shifted by $-1$)
are shown for $d=6$ and $7$ with $\ell=1$.
}
\end{figure}
\\
%
However, both $L_1/L_{1(GR)}$ and $L_2/L_{2(GR)}$ diverge in this limit;
at the same time, the ratio  $S_{d}/S_{d-1}$ approaches a constant nonzero value
($e.g.$ $S_{d}/S_{d-1}\simeq 0.14$, $0.078$ and $0.095$
for $d=5$, $6$ and $7$, respectively).
Moreover, for sufficiently large values of $Q$, 
the GR results are recovered.

For $d=5$, we have the choice to consider an electric or a magnetic charge on the brane.
However, similar to the case of  Einstein-Maxwell gravity,
one can show that
the properties of the solution are the same in both cases
and the electric-magnetic U(1)  duality still holds\footnote{We thank H. Reall
for a correction on this issue.}.
Our numerical results for $d=5$ are in good agreement 
with those in \cite{Kaus:2009cg}.
(We note, however, that in \cite{Kaus:2009cg} a different parametrization 
of the metric was employed.)

\subsubsection{The large black holes limit}
The observation that $\xi_0/\ell$ diverges as $a(0) \to a_c$,
implies that one can use the asymptotic expressions
(\ref{even-inf}) and (\ref{odd-inf})
to derive a number of simple analytic results for large black holes,
and thus to understand the large-$Q$ behaviour,
demonstrated in Figure 5.
Following Kaus and Reall \cite{Kaus:2009cg}, 
one starts by assuming an asymptotic expression for the parameter $U$ 
(which enters the large-$\xi$ expansion (\ref{even-inf}), (\ref{odd-inf}))
in terms of a power series in $\epsilon=\ell/\xi_0$.
Then the first junction condition (\ref{cond1}) implies that 
\begin{eqnarray}
\label{rb1}
U=\frac{1}{(d-4)^2}-\frac{(d-3)(4d-11)}{4(d-2)(d-4)^2}\epsilon^2+\dots~.
\end{eqnarray}
This shows that the metric function $a(\xi_0)$ 
($i.e.$, the size of the $AdS_2$ part of the brane metric) 
has the following approximate form on the brane
\begin{eqnarray}
\label{rb2}
a=L_1^2=\frac{1}{ (d-4)^2}\frac{\ell^2}{\epsilon^2}-\frac{(d-1)\ell^2}{4(d-2)(d-4)^2} +\dots~.
\end{eqnarray}
The second junction condition (\ref{cond1}) gives 
for the charge parameter the expression
\begin{eqnarray}
\label{rb3}
 Q^2=\frac{d-3}{2(d-4)^3}\frac{\ell^2}{\epsilon^2}-\frac{(d-3)(7d-18)}{8(d-2)(d-4)^3}\ell^2 +\dots~.
\end{eqnarray}
Inversion of eqs.~(\ref{rb1}) and (\ref{rb2}) 
yields $\epsilon$ as a function of $Q^2$.
One obtains
\begin{eqnarray}
\label{rb4}
&&
L_1^2=\frac{2(d-4)}{(d-3)}Q^2+\frac{6d-19}{4(d-2)(d-4)^2}\ell^2 +\dots,
\\
&&
L_2^2=\frac{2(d-4)^3}{(d-3)}Q^2+\frac{7d-18}{4(d-2)}\ell^2 +\dots~,
\end{eqnarray}
where comparison with (\ref{GR1}) (with $D=d-1$) 
shows agreement with GR in the limit of large $Q/\ell$. 
(Note that the leading order corrections are strictly positive.)

The same holds also for the entropy of these solutions, 
which, in $d>5$ dimensions, is
given by
\begin{eqnarray}
\label{arb3}
S_d= 2\frac{V_{d-3}}{4 G_d}\frac{\ell}{(d-3)}\xi_0^{d-3}
\left(
1-\frac{3(d-3)(d-4)}{2(d-2)(d-5)}\frac{\ell^2}{\xi_0^2}
\right)+\dots~.
\end{eqnarray}

The first term here corresponds to the entropy of an extremal RN black hole 
in $d-1$ dimensions. (Recall that $G_{d}=\frac{2\ell}{d-3} G_{d-1}$.)
With help of relation (\ref{rb3}), 
one can reexpress $S_d$ as a function of the charge parameter $Q$. 
As seen from the large-$Q$ region of Figure 4,
the analytical and numerical results are in excellent agreement.

For $d=5$, the leading order correction for the entropy as a function of $Q$
contains a $log$ term, which is absent in other dimensions. 
The corresponding relations  are given in \cite{Kaus:2009cg}, 
together with some relevant plots\footnote{For $d=5$, the
electric charge is ${\mathcal Q}=\frac{L_2^2}{L_1^2}Q$.
}.

\section{Further remarks }
With this work we have addressed the issue of black hole solutions
in the RS infinite braneworld scenario,
allowing the black holes to carry electric charge on the brane.
In the first part of the paper we have considered vacuum black holes
and nonextremal charged black holes on the brane.
Employing a different numerical technique
by solving the set of elliptic partial differential equations
in the full bulk, ranging from the brane to the AdS horizon,
we have obtained results that fully
support the claim by Yoshino \cite{Yoshino:2008rx} for the nonexistence of
static vacuum black holes
on an asymptotically flat brane in the RS infinite braneworld model.
This conclusion does not change when (nonextremal) solutions are considered,
which are charged with respect to a Maxwell field living on the brane.
Although 
``approximate'' solutions appear to exist for sufficiently small brane tension, 
these configurations are very likely only numerical artifacts. 

One should emphasize that, of course, 
the numerical results cannot be used to prove the nonexistence of 
static black hole solutions on the brane.
To clarify the issue of the existence of black holes on the brane 
one would thus need to either find a full analytic solution
or, alternatively, to provide a rigorous theoretical argument 
for the absence of such black hole solutions.
Lacking both, however, we think that a natural interpretation 
of our and previous numerical results is provided by the
conjecture that nonextremal braneworld black holes
would necessarily be time-dependent 
\cite{Emparan:2002px,Tanaka:2002rb}.
 
The situation is different for extremal black holes,
since the conjecture put forward by 
Emparan et al.~\cite{Emparan:2002px} and Tanaka \cite{Tanaka:2002rb}
does not forbid the existence of static extremal black holes.
Thus, in principle, the presence of a second global charge, apart from the mass,
could allow for the existence of extremal black holes also in a braneworld context.
While the numerical construction of localized extremal 
black holes still represents a numerical challenge to be met,
we here have considered the less ambitious task 
of constructing only the near-horizon geometry 
of extremal solutions with a Maxwell field on the brane.
In particular, 
we have found that the GR predictions in $d-1$ dimensions are 
reproduced for extremal black holes which are sufficiently large
as compared to the AdS scale. 

One should mention, though, that finding local solutions 
in the vicinity of the horizon does not guarantee the
existence of the corresponding global solutions.
(Chen et al.~\cite{Chen:2009rv}, for instance,
give an example where the physically relevant global solutions 
are absent despite the presence of a closed form near-horizon solution.) 
In our opinion, any progress in this direction 
would require the development of a consistent numerical scheme 
capable to achieve the explicit construction of the bulk extremal black holes.

However, another physically interesting situation to consider in the context 
of the RS infinite braneworld model concerns the case of 
static localized solitons on the brane.
With no event horizon present,
solitons do not possess intrinsic thermodynamical properties.
Thus the conjecture of Emparan et al.~\cite{Emparan:2002px}
and Tanaka \cite{Tanaka:2002rb}
has no obvious bearing on the existence of solitons on the brane.
 
Let us thus consider 
as the simplest example of (possible) solitons 
on the brane in the RS model,  
the case of brane world $Q$-balls
based on the matter field Lagrange density
\be                   \label{lQ}
{\mathcal L}_{Q} =  
\partial_\mu \Phi^\ast \partial^\mu \Phi  +U(|\Phi|).
\ee 
The matter field in this case is a complex scalar field $\Phi$ 
with a harmonic time dependence
and a non-renormalizable self-interaction
described by the potential $U(|\Phi|)$. 
The flat spacetime solutions of this theory 
were considered for the first time by Coleman \cite{Coleman:1985ki},
while their Einstein gravity generalizations were discussed in 
\cite{Lee:1991ax,Friedberg:1986tq,Kleihaus:2005me}.
These gravitating nontopological solitons describe localized particlelike
objects with finite energy.
 
We have attempted to construct $d=5$ spherically symmetric 
gravitating $Q$-balls
in the RS model, where the scalar field is confined to the brane
and described by the simple ansatz
\be                       
 \Phi=f(r)e^{-iw t},
\ee 
where we have followed a basically similar approach for these 
gravitating $Q$-balls
as for the static black hole solutions.
In particular, we have tried to solve the bulk Einstein equations 
numerically for the metric ansatz (\ref{metric-ansatz}) with $r_0=0$,
where the boundary conditions at $r=0$, $r=\infty$ and $\chi=0$
are still given by (\ref{bc1})-(\ref{bc3}) 
(with $r_0=0$) and (\ref{eqs-israeljc}), respectively,
and where the energy-momentum tensor is given by
\be
\label{tik1}
t_{\mu\nu}=\partial_\mu\Phi^\ast\partial_\nu\Phi+
\partial_\nu\Phi^\ast\partial_\mu\Phi-g_{\mu\nu}{\mathcal L}_{Q}.
\ee 

However, unexpectedly,
our attempts to construct gravitating $Q$-balls
on the brane have not been successful.
As in our attempts to construct static black hole solutions on the brane,
we have failed here to obtain reliable numerical 
gravitating $Q$-ball solutions on the brane.
Interestingly, the reason for the numerical inaccuracy of the configurations
and the lack of the numerical convergence again resides in
the unnatural far field behaviour of the metric functions
close to the AdS horizon.
In fact, this unnatural far field behaviour 
is completely analogous to one observed 
for the configurations supposed to describe static black holes.
 
Clearly, this behaviour is hard to understand,
since there is no clear a priori reason for solitons {\sl not} to exist on the brane.
However, if we interpret our results for the static (nonextremal)
black hole solutions as implying, that there are no such solutions
on the brane (satisfying the given set of equations and boundary conditions),
we must - consistently - draw the same conclusion for the gravitating $Q$-ball
solutions.
While it might be interesting to
attempt the construction of other types of solitons on the brane,
the present results strongly discourage such a construction,
since the same problems in the vicinity of the AdS horizon are likely
to arise.

How do we then judge the issue of black holes and regular solutions
on the brane? There are still the positive results from the near-horizon
black hole solutions in the extremal case,
but these extremal near-horizon solutions avoid the problematic region
in the vicinity of the AdS horizon. 
Thus a full calculation of
extremal solutions is clearly called for, where one attempts
to smoothly extend the near-horizon geometry 
of the extremal solutions into the asymptotic region,
in spite of the tremendous numerical challenge.
The final outcome might, however, be that one is led to conclude,
that there are no extremal black hole solutions on the brane, either.

\section*{Acknowledgements}
We thank H. Reall for important comments on a first version of this work.
E.R.~would like to thank M.~Volkov for useful discussions.
B.K.~gratefully acknowledges support by the DFG. 
The work of E.R.~was supported by a fellowship 
from the Alexander von Humboldt Foundation and the
Science Foundation Ireland (SFI) project RFP07-330PHY.

\appendix
\section{The numerical method}

The set of three coupled non-linear
elliptic partial differential equations
for the functions $A$, $B$ and $C$
has been solved numerically for the $(\rho,\chi)$ coordinate system 
introduced in Section 2.1.1,
subject to the boundary conditions (\ref{bc1})-(\ref{bc-brane2}).

The first step has been to introduce a new radial variable  
$x=\rho/(1+\rho)$ 
which maps the semi infinite region $[0,\infty)$ to the finite region $[0,1]$.
This involves the following substitutions in the differential equations
\begin{eqnarray}
\rho u_{,\rho}   \longrightarrow    (1-x) f_{,x},
~~~
\rho^2   u_{,\rho \rho}   \longrightarrow
(1-x)^2    u_{,xx}
  - 2 (1-x) u_{,x}
\end{eqnarray}
where $u$ denotes any of the unknown functions $A$, $B$, $C$.

Next the equations for these functions
are discretized on a non-equidistant grid in $x$ and $\chi$. 
Here we have considered various grid choices,  
the number of grid points ranging between $180 \times 30$ and $80 \times 70$. 
The grid covers the integration region
$0\leq x \leq 1$ and $0\leq \chi \leq \pi/2$.  
Typically, the mesh in the $x-$direction has been
denser in the near-horizon region ($x=0$)
and for values of $x$ close to the AdS horizon.
Most of the $\chi$-meshes employed have been equidistant.

All numerical calculations have been 
performed by using the programs FIDISOL/CADSOL \cite{schoen}.
Here we shall briefly review its basic aspects.
The code requests the system of nonlinear partial differential equations 
in the form
\begin{eqnarray}
P(x,y,u,u_{x},u_{y}
,u_{x y},u_{xx},u_{yy})=0,
\end{eqnarray}
subject to a set of boundary conditions on a rectangular domain.
(For convenience, we have used the notation $\chi\equiv y$.)
Besides the set of equations, FIDISOL/CADSOL requires
the boundary conditions, the Jacobian matrices for the equations
and the boundary conditions, as well as some initial guess for the functions.
The Jacobian matrices are generated by simple differentiation of each equation $P$
with respect to $u,u_{x},u_{y}
,u_{x y},u_{xx}$ and $u_{yy}$.

FIDISOL/CADSOL uses a Newton-Raphson method.  
The numerical procedure works as follows:
for an approximate solution $u^{(1)}$,
$P(u^{(1)})$ does not vanish.
The next step is then to consider an improved  solution
\begin{eqnarray}
u^{(2)}=u^{(1)}+w \Delta u,
\end{eqnarray}
supposing that $P(u^{(1)}+w \Delta u)=0$ 
(with $w$ a relaxation factor, which is usually chosen as $w=1$).
The expansion in the small parameter
$\Delta u$ gives to first order
\begin{eqnarray}
\label{new1}
0=P(u^{(1)}+\Delta u) \approx
P(u^{(1)})+\frac{\partial P}{\partial u }(u^{(1)}) \Delta u \ +\dots \ .
\end{eqnarray}
This equation is then used to determine the correction
$\Delta u^{(1)}= \Delta u$.
Repeating these calculations iteratively 
($u^{(3)}=u^{(2)}+\Delta u$ 
etc.), the approximate solutions will converge,
provided the initial guess is close enough to the true solution.
In each step, a linear system of equations is solved, and
 the residual $||P(u^{(i)})||$
decreases by a factor of approximately $10-20$.
The iteration stops after $i$ steps, when the Newton residual $P(u^{(i)})$
is smaller than a prescribed tolerance.
Clearly, it is essential to have a good first guess, 
to start the iteration procedure.

The package FIDISOL/CADSOL provides also error estimates for each function,
which allows to judge the quality of the computed solution.
The errors are computed on the ``consistency level'', namely, 
the discretized Newton residual,
and as discretization error terms in $x,y$ directions.
The discretization error is estimated through the difference of
difference quotients. 
For example, in (\ref{new1}), the derivative of the solution $u$ 
and of the correction function $\Delta u$
are discretized by a difference method with arbitrary consistency orders.
Derivatives are replaced, for example, in the form 
$u_{xx}\Leftarrow u_{xx,d}+d_{xx}$,
$\Delta u_{xx}\Leftarrow \Delta u_{xx,d}$,  
where the index $d$ means ``discretized''.
$d_{xx}$ is the estimate for the discretization (or truncation) error 
of $u_{xx}$, defined as $d_{xx}=u_{xx,d,next}-u_{xx,d}$, 
where the index ``$next$'' denotes the next higher member of the
family of nonequidistant backward difference formulas.
The discretized Newton residual decreases with the number of 
Newton-Raphson iterations, while
the discretization error terms depend on the grid size 
and the used consistency order, 
$i.e.$ on the order of the discretisation of derivatives (in our work,
this order was six).
Furthermore, the error terms are used for 
the determination of stopping criteria for the Newton-Raphson method.
Further details on the numerical method and explicit examples are provided in
\cite{schoen}.

In this scheme, the input parameters are the event horizon radius $r_0$
and the value $\ell$ of the AdS length scale 
which form the dimensionless parameter $L=r_0/\ell$.
In our approach we set $r_0=1$
and we start with the Schwarzschild-Tangherlini solution 
as initial guess ($i.e.$, $L=0$ and $A=B=C=0$).
Then we increase the value of $L$ slowly.
The iterations converge, and, in principle, 
repeating the procedure we obtain
in this way solutions for higher values of $L$. 
In some of the calculations, we interpolate the resulting
configurations on points between the chosen grid points,
and then use these for a new guess on a finer grid.

 \begin{small}
 
 \end{small}
\end{document}